\documentclass{article}

\PassOptionsToPackage{numbers, compress}{natbib}
\usepackage[preprint]{neurips_2024}

\usepackage[utf8]{inputenc} 
\usepackage[T1]{fontenc}    
\usepackage{hyperref}       
\usepackage{url}            
\usepackage{booktabs}       
\usepackage{amsfonts}       
\usepackage{nicefrac}       
\usepackage{microtype}      
\usepackage{xcolor}         
\usepackage{amssymb} 
\usepackage{amsmath}
\usepackage{amsthm}
\usepackage{enumitem}
\usepackage{algorithm}
\usepackage{algorithmic}
\usepackage{listings}
\usepackage{graphicx}
\usepackage{caption}
\usepackage{subcaption}

\DeclareMathOperator*{\argmax}{arg\,max}

\title{Responding to Promises: No-regret learning against followers with memory} 

\newtheorem*{thm*}{Theorem} 
\newtheorem{thm}{Theorem}
\newtheorem{lem}[thm]{Lemma}

\newtheorem{prop}[thm]{Proposition}
\newtheorem{claim}[thm]{Claim}

\newtheorem{coro}[thm]{Corollary}

\def\ddefloop#1{\ifx\ddefloop#1\else\ddef{#1}\expandafter\ddefloop\fi}

    \def\ddef#1{\expandafter\def\csname c#1\endcsname{\ensuremath{\mathcal{#1}}}}
    \ddefloop ABCDEFGHIJKLMNOPQRSTUVWXYZ\ddefloop

    \def\ddef#1{\expandafter\def\csname s#1\endcsname{\ensuremath{\mathsf{#1}}}}
    \ddefloop ABCDEFGHIJKLMNOPQRSTUVWXYZ\ddefloop

    \def\ddef#1{\expandafter\def\csname b#1\endcsname{\ensuremath{\mathbb{#1}}}}
    \ddefloop ABCDEFGHIJKLMNOPQRSTUVWXYZ\ddefloop
    
    \def\bbDelta{\boldsymbol{\Delta}} 

\author{%
  Vijeth Hebbar 
  \\
  Coordinated Science Laboratory \\
  University of Illinois Urbana–Champaign\\
  Urbana, IL 61801, USA. \\
  \texttt{langbort@illinois.edu} 
  \And
  C\'edric Langbort 
  \\
  Coordinated Science Laboratory \\
  University of Illinois Urbana–Champaign\\
  Urbana, IL 61801, USA. \\
  \texttt{vhebbar2@illinois.edu} 
}

\begin{document}

\maketitle

\begin{abstract}
    We consider a repeated Stackelberg game setup where the leader faces a sequence of followers of unknown types and must learn what commitments to make. While previous works have considered followers that best respond to the commitment announced by the leader in every round, we relax this setup in two ways. Motivated by natural scenarios where the leader's \emph{reputation} factors into how the followers choose their response,  we consider followers with memory. Specifically, we model followers that base their response on not just the leader's current commitment but on an aggregate of their past commitments. In developing learning strategies that the leader can employ against such followers, we make the second relaxation and assume boundedly rational followers. In particular, we focus on followers employing \emph{quantal responses}. Interestingly, we observe that the smoothness property offered by the quantal response (QR) model helps in addressing the challenge posed by learning against followers with memory. Utilizing techniques from online learning, we develop algorithms that guarantee $O(\sqrt{T})$ regret for quantal responding memory-less followers and $O(\sqrt{BT})$ regret for followers with bounded memory of length $B$ with both scaling polynomially in game parameters. 

  
\end{abstract}





\section{Introduction}





Stackelberg games (SGs) offer a natural framework to capture strategic interactions with hierarchical play between agents. They have been used to model scenarios in a wide range of domains like defence \cite{tambe2011security}, market behaviour \cite{anderson1992stackelberg}, persuasive signaling \cite{kamenica2011bayesian,hebbar2020stackelberg,massicot2019public} and more \cite{farokhi2016estimation, zrnic2021leads}. This hierarchy is typically delineated by naming one agent as the \emph{leader} (she), who acts first and commits to a strategy, and the other as a \emph{follower} (he), who then best responds to the leader's committed strategy. A key feature of SGs is the leader's ability to estimate the follower's best response, without which she cannot compute the optimal strategy to commit to. However, in the event that the follower's payoff function is unknown, a possible (and well-studied) work-around for the leader is to learn her  optimal strategy via repeated interaction with the follower, resulting in a so-called repeated SG \cite{letchford2009learning,blum2014learning,balcan2015commitment,sessa2020learning,castiglioni2020online}. In this work, we consider one such setup where a leader is facing a sequence of followers whose type (and hence, payoff function) is unknown to her and is tasked with finding her optimal strategy.

When learning in such repeated games, the leader sequentially updates her strategy in every round of the repeated game based on feedback from past interactions with the followers. In the event that the leader's strategy in revealed to the follower in each round,\footnote{Such revelation can take various forms, e.g. through attacker surveillance in security games \cite{tambe2011security} or leader announcements in persuasive signaling \cite{kamenica2011bayesian,massicot2019public}} a prevalent assumption in the literature is that the follower selects his best response to this strategy \cite{balcan2015commitment,castiglioni2020online}. While this assumption is well justified in single-shot SGs or in repeated SGs where the leader consistently plays a single strategy, it may not always hold when she continually changes her strategy.  Indeed, one potential reason for this deviation -- one that becomes the focus of our study -- could be that the follower possesses memory and bases decisions not only on the leader's strategy in the current round,  but also on her past plays. 
This could occur, e.g., in scenarios where the leader merely \textit{`announces'} her committed strategy\footnote{Readers are referred to the discussion by \citet{kamenica2011bayesian} for more details on leader commitment in such settings.} (e.g. persuasive signaling  \cite{kamenica2011bayesian,massicot2019public}) and does not `act' it out (like, e.g., actually deploying patrols in security games \cite{tambe2011security}) before the follower picks his response. In such cases, the follower de facto responds to a promise from the leader (hence the title of this paper) and, if the veracity of this promise can only be verified ex-post, the reputation or credibility of the leader become paramount. Using an average of past promises (or any other aggregation of past plays) is a way for a follower endowed with memory to generate and evaluate such a reputation.


Note that we are not claiming that it is necessarily beneficial for the follower to act in this way. 
We are, however, saying that such reliance on past promises should be considered to be within the range of possible (or even expected) behaviors of a boundedly rational follower. Consider, e.g., a politician interacting with a diverse set of voters over the course of a public campaign. Even if the politician intends to honor the promises she makes, a skeptical voter will base their voting decisions on her past promises as well and not solely the one she made to him. 

In the presence of such followers, a very natural question arises for the leader, namely:

\textbf{Q:} How can a leader learn her optimal strategy when facing a sequence of followers that make decisions that depend on a history of her strategies? 

In this work, we will focus solely on a process whereby the followers respond to a weighted average of past leader plays (which we denote as ``reputation'', in accordance with the discussion above). 

\subsection{Road-map and Contributions}
In Section \ref{sec:OL-prelim}, we present some background on online optimization that is relevant to our work. We then formalize our problem framework in Section \ref{sec:setup}, focusing on a matrix Stackelberg game played between a leader and a sequence of followers of unknown type. First, we consider memory-less followers that base their decisions only on the leader's current strategy in every round. However, we introduce a generalized response model for the followers, representing a departure from previous approaches. This model enables us to accommodate quantal and best-responding followers, among others. Subsequently, we extend our framework by endowing followers with memory, marking the second point of deviation from existing literature. 

We then present our results in Section \ref{sec:results} in the form of two algorithmic approaches designed to sequentially pick leader strategies. Our first approach tackles learning against memory-less follower and guarantees $\cO(\sqrt{T})$ regret (Theorem \ref{thm:main-regret-bd-no-mem}). This extends existing results \cite{balcan2015commitment,castiglioni2020online} to scenarios where followers may employ quantal responses instead of best responses. Our second approach targets the problem of learning against followers with memory and once again guarantees $\cO(\sqrt{T})$ regret (Theorem \ref{thm:main-regret-bd-w-mem}). This alternate approach focuses on quantal responding followers and relies crucially on the smoothness properties of such a response model. Thus, in addition to the QR model's ability to capture more realistic follower behaviour, our result showcases its usefulness in enabling leader's learning.  Finally, we discuss the computational challenges involved in both these approaches and acknowledge that, akin to related work \cite{balcan2015commitment,sessa2020learning}, our approach is also computationally expensive. 



\subsection{Related Work}

\paragraph{Learning in Stackelberg Games} Research on learning optimal Stackelberg strategies from repeated Stackelberg games (SGs) has explored various scenarios where the leader has  incomplete information about the follower's decision-making process. 
Several works consider a leader repeatedly interacting with a single follower and use best response queries to learn the optimal Stackelberg strategy \cite{letchford2009learning,marecki2012playing,blum1997universal}.  In contrast, in our work we consider a leader that faces a (possibly, adversarially generated) sequence of followers of unknown \emph{types}. Such a setup was first considered by \citet{balcan2015commitment}, who frame the problem as an instance of online learning and develop no-regret learning algorithms for it. However, a key assumption in theirs and other similar studies \cite{castiglioni2020online,sessa2020learning,velicheti2024learning} is that at every time step the followers best respond to the leader's current commitment in that step. 
In a bid to capture reputation effects, we relax this assumption, and allow the followers' responses to depend on the leader's past decisions (albeit in a structured way). Notably, \citet{xu2016playing} consider a Stackelberg security game (SSG) framework \cite{tambe2011security} and develop an online learning algorithm that is agnostic to nature of attackers' (followers') response. However, their approach exploits the structural properties of SSGs and it is unclear how their approach could be extended to general Stackelberg matrix games. 

There are multiple lines of research that study variants of repeated SGs where followers are not best responding. A widely studied setup considers followers to also be learning agents that rely on their past payoffs to make decisions rather than the leader's strategy \cite{zrnic2021leads,lin2024persuading,deng2019strategizing,cesa2013online,fiez2020implicit}. \citet{haghtalab2024calibrated} consider learning from a single follower that, in the absence of access to leader's strategy, responds to a calibrated forecasts of it. In contrast to these works, we assume that followers have access to the leader's commitments in every round. Another work considers learning against a single \emph{non-myopic} follower that strategically deviates from best-responding to the leader's strategy to mislead the leader's learning process \cite{haghtalab2022learning}. However, in our research, followers deviate from the best response not to mislead, but because they also consider the leader's past strategies when making decisions.


\paragraph{Quantal Stackelberg Games} Quantal Reponse (QR) is a popular choice for modelling bounded rationality of agents and has been used to model human behavior in various game theoretic settings \cite{mckelvey1995quantal, feng2024rationality,yang2012computing}. Beyond its practical significance, the mathematical structure of the underlying logit choice model makes it a ubiquitous presence in applications like reinforcement learning \cite{sutton2018reinforcement}, neural networks \cite{goodfellow2016deep} and in online learning \cite{arora2012multiplicative}. Exploiting this mathematical structure, \citet{wu2022inverse} demonstrate that the smoothness property of QR models improve the query efficiency in learning a single follower's payoff functions. 
In a similar spirit, we show that when facing a sequence of memory-endowed followers, this smoothness property allows us to develop no-regret algorithms.     


\paragraph{Online Learning with Memory}
Our work is also closely related to the problem of online learning against adversaries with memory \cite{arora2012online,anava2015online,hebbar2023online} i.e. online optimization setups where the stage costs are allowed to depend on past learner decisions. Indeed, we observe that when follower's have memory our problem can be cast as an instance of such an online optimization problem. However, non-convexity of stage costs in our framework possess the additional challenge. To address this, we extend approaches developed for online non-convex learning with \emph{memory-less} costs \cite{agarwal2019learning,suggala2020online} to our problem.   




\section{Preliminaries on online learning} \label{sec:OL-prelim}


In this section, we present a quick summary of concepts in online optimization that are relevant to our work. Let $\{f^t\}_{t=1}^H$ denote the sequence of cost functions -- mapping decision set $\cX \subset \bR^n$ to $\bR$ -- faced by the learner over a horizon of length $H$. In line with the standard online optimization framework, we assume that when making decision $x^{t}$ the learner only has access to $\{f^\tau\}_{\tau=1}^{t-1}$. The learner wishes to pick decisions $\{x^t\}_{t=1}^H$ such that their regret, defined as
    \begin{flalign} \label{eq:regret-OL}
        \cR(H) = \sum_{t=1}^H f^t(x^t)  - \min_{x\in \cX} \sum_{t=1}^H f^t(x),
    \end{flalign}
 grows sub-linearly with $H$.
Let us now look more closely at one class of algorithms that are routinely applied to solve online optimization problems: Follow-The-Perturbed-Leader (FTPL) \cite{kalai2005efficient} (or equivalently, Follow-The-Regularized-Leader (FTRL) with randomized regularizers \cite{hazan2016introduction}). Let us suppose that we have access to an approximate optimization oracle $\bO_\epsilon$ which,  given $f$ and set $\cX$, outputs 
\begin{equation}
    x^* = \bO_\epsilon\big(f,\cX\big) \quad \mbox{ such that } \quad f(x^*) \leq \min_{x\in \cX} f(x) + \epsilon. \notag 
\end{equation}
Now, given a random vector $\sigma\in \bR^m$ picked from distribution $\cD$ and a regularizing function $\rho:\bR^m \times \cX \to \bR$, suppose that we generate decisions $\{x^t\}_{t=1}^T$ according to
\begin{flalign} 
    x^1 = \bO_{\epsilon}\big(\rho(\sigma,\cdot),\cX \big) 
    \text{ and } x^{t} = \bO_{\epsilon}\bigg(\sum_{\tau=1}^{t-1} f^\tau+\rho(\sigma,\cdot) , \cX\bigg), \; \forall t>1 . \label{eq:approx-FTPL}
\end{flalign}
Note that the decisions $\{x^t\}_t$ generated above are themselves random variables due to their dependence on $\sigma$ and so, we are interested in bounding the expected value of the regret defined in (\ref{eq:regret-OL}). We can now state the following refinement of the well known bound from \cite{kalai2005efficient,cesa2006prediction} on the regret of FTRL algorithms.

\begin{thm} \label{thm:FTPL-general-bound}
    When $\{x^t\}_t$ is generated according to (\ref{eq:approx-FTPL}) we can bound the expected regret as
    \[
        \bE[\cR(H)] \leq \sum_{t=1}^H \bE[f^t(x^t)-f^t(x^{t+1})] + \bE[\rho(\sigma,x')-\rho(\sigma,x^1)] + \epsilon (H+1) 
    \]   
\end{thm}
The terms of the form $\bE[f^t(x^t)-f^t(x^{t+1})]$ are typically called \emph{stability} error terms and bounding them is a standard step in the regret analysis of many online learning algorithms. As we see ahead, that will be the case for our approach as well. Finally, we state a proposition that will aid us with bounding these stability error terms in our setup. 
\begin{prop}(\citet{kalai2005efficient}) \\ \label{prop:diff_exp_value} 
    Let $f$ be a bounded integrable 
    function mapping $\bR^n \to  \bR$. Let $\sigma \sim U[0,\frac{2}{\nu}]^n$ and $|f| \leq B$ for some constants $B,\nu$. Then for any $G,c\in \bR^n$  
    \begin{flalign*}
        \big|\bE[f(G+c+\sigma) - f(G+\sigma)]\big| \leq \nu B \|c\|_1   
    \end{flalign*}
\end{prop}


\section{Problem Setup} \label{sec:setup}

Before we begin describing our framework, let us define some notation to aid with exposition in the remainder of this paper. For any integer $S\in \bZ_{\geq 0}$, we define the set $[S] \triangleq \{1,\dots,S\}$. We denote the $j^{th}$ column of a matrix $A$ as $A_j$ and the element corresponding to the $i^{th}$ row and $j^{th}$ column as $A_{ij}$. The maximal absolute element of a matrix $A$ is denoted by $\overline{A} \triangleq \max_{i,j} |A_{ij}| $. We denote the induced $p$-norm of a matrix $A$ as $\|A\|_p\triangleq sup_{\|x\|_p=1} \|Ax\|_p$.  
For an integer $n\in \bZ_{\geq 0}$ we denote the standard $n$-dimensional simplex as $\bbDelta_n$ i.e. $\bbDelta_n = \{s\in \bR^n_{\geq 0}|\sum_i s_i = 1\}$.  Equipped with this notation, let us now formally set-up our problem.

\subsection{Single-shot Stackelberg game}
Consider a Stackelberg matrix game between a single leader (she) and a follower (he) that has one of $K$ types. The leader's utility matrix is $U \in \bR^{N\times M}_{\geq 0}$, while a follower of type $k \in \{1,\dots,K\}$ has utility matrix  $V^k\in \bR^{N\times M} $. Here, $N$ and $M$ are the number of actions for the leader and follower respectively. We can represent any strategy played by the leader (resp. follower) as an element $x$ (resp. $y$) of $\bbDelta_N$ (resp. $\bbDelta_M$) where $x_i$ (resp. $y_i$) denotes the probability of them playing action $i$. Then, the expected payoff of the leader and follower (of type $k$) can be captured simply as $x^\intercal U y$ and $x^\intercal V^k y$ respectively. 

Let us now consider a generalized model of a \emph{memoryless} follower's response, focusing for now on type $k$ . Given a leader strategy $x\in \bbDelta_N$, we define the follower's \emph{perturbed} response function as 
\begin{equation}
    y^k(x) \in \argmax_{y\in \bbDelta_M} x^\intercal V^k y  + \frac{1}{\eta} h(y) .  \label{eq:gen-fol-BR} 
\end{equation}
where $h:\bbDelta_M \to \bR$ is a perturbation function. Various choices of perturbation functions can be made to model a variety of decision-making behaviour \cite{fudenberg1998theory}. 
However, for the purpose of our work, we are interested in two primary choices: one when $h = 0$, modelling a perfectly rational follower and the other, when it is the Gibbs entropy function $h(y) = -\sum_{i=1}^m y_i \ln y_i$, modelling a quantal responding follower \cite{mertikopoulos2016learning}. To highlight their nature, we will denote the follower's response in the absence of perturbation (i.e. $h=0$) as $y_{BR}^k(x)$ and under the entropy perturbation function as $y_{QR}^k(x)$. 


It is important to note here that $y_{BR}^k(x)$ may not be uniquely defined for every value of $x$. While we re-visit this issue in later sections, for now we assume that all players break ties according to some 
common deterministic rule. On the other hand, $y_{QR}^k(x)$ is uniquely defined as it is the solution to a convex optimization problem. In-fact, it is well known \cite{mertikopoulos2016learning,arora2012multiplicative,sutton2018reinforcement} that it takes the closed form given by the logit choice model 
\[y^k_{i,QR} = \frac{\exp{\big(\eta \langle V^k_i, x \rangle\big)}}{\sum_{j=1}^m \exp{\big(\eta \langle V^k_j, x \rangle \big)}} \quad \forall i\in[M].\]

In the context of the QR model, $\eta$ acts as the bounded rationality constant and for the purpose of our work, we will assume that it is known and constant across all follower types. The QR model also offers some nice smoothness properties for the follower's response, namely we can make the following claim: 

\begin{claim} \textbf{(Lipschitz Quantal Response)}
\label{claim:diff_y_bd}
    For any $x,z\in \bbDelta_N$ and $k\in [K]$ we have
    \[\|y^k_{QR}(x) - y^k_{QR}(z)\|_1 \leq L \|x-z\|_\infty \] 
    where $L= 2 \eta \max_{k} \|V^k\|_1$ .
\end{claim}

All proofs can be found in Appendix \ref{app:proofs}. 
To end our discussion on the follower's response model, we define the matrix valued function $\cY:\bbDelta_N \to \bR^{M\times K}$ by stacking the response functions from (\ref{eq:gen-fol-BR}) for each follower type as \[\cY(x) \triangleq [y^1(x),\;\dots, \; y^k(x),\; \dots, \; y^K(x)].\] As before, when we wish to highlight the nature of the response function, we will use $\cY_{BR}(x)$  and $\cY_{QR}(x)$ for best responding and quantal responding follower's respectively. Let $e^k$ denote the $k^{th}$ basis vector in $\bR^K$. Then, note that we can recover the response of a follower of type $k$ simple as $y^{k}(x) = \cY(x) e_k$. 

Finally, let us characterize the leader's utility when faced with a single follower of type $k$. For now, we do not impose any specific form on the follower's response function $y^k(\cdot)$, so long as it follows the general structure outlined in (\ref{eq:gen-fol-BR}) and is uniquely defined. However, we assume that the leader is informed about the follower's response mechanism, allowing her to evaluate her payoff function as $x^\intercal U \cY(x) e_k$ for every strategy $x\in \bbDelta_N$. Notably, this assumption implicitly means that the leader knows the follower payoff matrices $\{V^k | k\in [K]\}$ without which she cannot evaluate $\cY(\cdot)$. When a follower is quantal responding, we can extend the smoothness property presented in Claim \ref{claim:diff_y_bd} to the leader's payoff function and state
\begin{coro} \label{cor:Lip_leader_util}\textbf{(Lipschitz continuity of leader utility.)}
    Let $g = [g_1, \dots, g_K] \in \bR^K_{\geq 0}$. Then
    \begin{flalign}
     |x^\intercal U \cY_{QR}(x) g - z^\intercal U \cY_{QR}(z) g| & \leq L_u \|x - z\|_\infty \notag \\
     \big|x^\intercal U \big(\cY_{QR}(x) - \cY_{QR}(z)\big) g\big| & \leq L_u' \|x - z\|_\infty \label{eq:aux-leader-util-bd}
     \end{flalign}
    where $L_u =  (1+L) \|g\|_1 \| U \|_1$ and $L_u' = L \| U \|_1 \|g\|_1$. 
\end{coro} 


\subsection{Repeated SG against a sequence of followers}
Consider now a repeated SG being played between the leader and a sequence of followers over a horizon of length $H$. Let $\{x^t\}_{t=1}^H$ be the sequence of the strategies that the leader commits to over the horizon $H$.  We represent the types of followers encountered by the leader through the sequence $\{g^t\}_{t=1}^H$, where $g^t \in \{e_k|k\in [K]\}$ at any time $t$, and $g^t=e_k$ signifies that the leader is facing a follower of type $k$ at $t$. 

Crucially, we will now assume that the leader is initially unaware of the sequence of follower types she will encounter and must learn what strategies to pick at every round based on past outcomes. While we will allow the sequence of types to be generated adversarially, we will consider the adversary to be \emph{oblivious} \cite{cesa2006prediction,hazan2016introduction}, i.e., the sequence of followers is determined without seeing any learner strategies. 

For the leader to have any meaningful chance to ``learn" what strategies to play, it is important for her to receive feedback. Two types of feedback are typically considered in such scenarios \cite{balcan2015commitment,castiglioni2020online}: \emph{full information feedback}, where the follower's type $g^t$ is revealed after the leader picks $x^t$ and \emph{partial information feedback}, where only the follower's response $\cY(x^t) g^t$ is revealed. For the purpose of this work, we will assume that leader's have full information feedback and leave the extension to other forms of feedback as future work. We can now consider followers of two types.



\subsubsection{Memory-less followers} \label{sec:follower-no-mem}
We say a follower is ``memory-less" when, at time $t$, his best response is $\cY(x^t)$, thus being independent of all past play by the leader. In this case, the cumulative payoff gained by the leader at the end of $H$ time steps 
is \(\sum_{t=1}^H \langle \cY(x^t)^\intercal U^\intercal x^t, g^t \rangle.\)
Naturally, the leader aims to achieve a high cumulative payoff, and if she had access to $g^t$ when picking her strategy $x^t$, she could simply pick $x^t$ as a maximizer of $x U \cY({x}) g^t$ at time $t$.

However, under our assumption that the type is revealed to her only after she picks $x^t$, the leader's problem turns into an instance of online learning \cite{hazan2016introduction,cesa2006prediction} with non-convex cost functions. Motivated by this observation, a natural measure of performance for the leader's decision making approach is \emph{regret}, defined for our problem as 
\begin{equation}
    Regret(H)\triangleq \max_{x\in \bbDelta_N} \langle \cY(x)^\intercal U^\intercal x, G^H\rangle - \sum_{t=1}^H \langle \cY(x^t)^\intercal U^\intercal x^t, g^t \rangle \quad \text{where $G^t \triangleq \sum_{\tau=1}^t g^\tau$}.
    \label{eq:regret-no-memory}
\end{equation}
In other words, we compare the cumulative payoff accumulated by the leader at the end of $H$ rounds to the payoff of her \emph{best-in-hindsight} strategy i.e. the static strategy that would have obtained her the highest total payoff. Then, our first goal is

\textbf{P1:} \textit{Design algorithms that sequentially pick leader decisions $\{x^t\}_{t=1}^T$ such that she has \emph{no-regret} i.e  $\frac{Regret(H)}{H}$ vanishes as $H$ goes to infinity.}

Notably, the presented setup with memory-less followers is closely related to the one considered by \citet{balcan2015commitment}. Indeed, if we further restrict the followers to be perfectly rational and replace the general response function $\cY$ with $\cY_{BR}$ it is identical to their setup. We will now introduce the primary point of deviation of our work and consider followers with memory.  

\subsubsection{Followers with memory} \label{sec:followers-w-mem}
We will now consider the situation where the followers base their action not only on the leader's current commitment, but also on her past commitments. While such dependence on the past can occur in many different ways, with the motivation of capturing the effect of leader's reputation on a follower's response, we assume that he bases his decision \emph{solely} on a weighted average of her past play. Formally, we define the \emph{time-averaged} leader strategy as 
\begin{equation} 
    z^{t} \triangleq \frac{1}{b_t} \sum_{\tau=1}^{t} a_{t-\tau} x^\tau \text{ where $b_t \triangleq \sum_{\tau=1}^{t} a_{t-\tau}$ and $a_s \in \bR_{\geq 0}$ $\forall \; s\geq 0$.} \label{eq:avg_leader-strat}
\end{equation}
Viewing $z^t$ as the \emph{reputation} established by the leader in the eyes of each follower through her past play, $a_{s}$ captures the weight of her decision from $s$ rounds ago on her current reputation. Such weighted averaging approach can capture a wide range of models of the follower's memory. For example, setting $a_s = 1$ when $s<B$ and $a_s = 0$ otherwise, models a follower with bounded memory of length $B$ who views reputation as a simple unweighted average of leader's play. Setting $a_s = \gamma^s$ for some $0<\gamma<1$ instead, captures a follower that models reputation as a discounted average of leader strategies. 

Regardless of the weights in (\ref{eq:avg_leader-strat}), we will assume that a follower's response at time $t$ is captured in $\cY(z^t)$. In other words, we model each follower as responding to an average of past play by the leader. Under this modelling assumption, the expected cumulative payoff of the leader at the end of $H$ rounds is simply \(\sum_{t=1}^H \langle \cY(z^t)^\intercal U^\intercal x^t, g^t \rangle \), which now depends explicitly on the time-averaged leader strategy $z^t$.   

Apart from capturing reputation effects, considering that each follower responds to such a weighted average has another important consequence.  To see it, consider a leader that adheres to a single strategy $x$ over all rounds of the repeated game. Owing to the averaging process, we then have $z^t = x$ for all $t \in [H]$ and consequently, the leader's total payoff is then $\langle \cY(x)^\intercal U^\intercal x, g^G \rangle$. Motivated by this, we can once again compares the leader's cumulative payoff to the payoff of the best \emph{static} action in hindsight and define regret defined as 
\begin{equation}
    Regret_M(H)\triangleq \max_{x\in \bbDelta_N} \langle \cY^\intercal(x) U^\intercal x, G^H\rangle - \sum_{t=1}^H \langle \cY(z^t)^\intercal U^\intercal x^t, g^t \rangle
    \label{eq:regret-memory}
\end{equation}
where the subscript $M$ highlights that we are considering followers with memory. Notably, in the absence of the averaging process in (\ref{eq:avg_leader-strat}), it can be challenging to define meaningful notions of regret when payoffs depend on past decisions in online optimization \cite{arora2012online,hebbar2023online}. We can now state our second goal as

\textbf{P2:} \textit{Design an algorithm that picks leader decisions $\{x^t\}_{t=1}^T$ such that she has \emph{no-regret} when facing followers with memory i.e  $\frac{Regret_M(H)}{H}$ vanishes as $H$ goes to infinity.}

In comparing with the regret definition in (\ref{eq:regret-no-memory}), we see that the benchmark to which we compare the leader's cumulative payoffs is identical in both the cases (with and without memory). As a result, developing algorithms that are no-regret w.r.t (\ref{eq:regret-memory}), means that despite the followers possessing memory, in the long run we are able achieve the performance of the best static action when followers are memory-less.



\section{Main Results} \label{sec:results}

\subsection{Learning with memory-less followers} \label{sec:result-no-memory}
Let us first consider the problem \textbf{P1} posed in Section \ref{sec:follower-no-mem}. We can state the following 
\begin{thm} \label{thm:main-regret-bd-no-mem}
    Let $\sigma = [\sigma_1,\dots,\sigma_K]$ denote a uniform random vector such that $\sigma_k \stackrel{i.i.d}{\sim} U[0,\nicefrac{2}{\nu}]$ for all $k\in [K]$. Then, setting $\nu = \sqrt{\nicefrac{K}{H}}$ and picking decisions according to (\ref{eq:approx-FTPL}) with $\cX = \bbDelta_N$, $\rho(\sigma,x) \triangleq - \langle \cY(x)^\intercal U^\intercal x, \sigma\rangle$ and $f^t = -\langle \cY(x)^\intercal U^\intercal x, g^t \rangle$ 
  gives us the following bound: 
    \begin{flalign*}
        \bE[Regret(H)] \leq 2 \overline{U} \sqrt{KH} + \epsilon (H+1)
    \end{flalign*}
    where the expectation is taken over the distribution of $\sigma$.
\end{thm}
\textbf{Proof Sketch:} We begin by invoking Theorem \ref{thm:FTPL-general-bound} which gives us 
\begin{flalign*}
         \bE\big[Regret(H)\big] \leq & \sum_{t=1}^H \underbrace{\bE\big[\langle \cY^\intercal(x^{t+1}) U^\intercal x^{t+1} - \cY^\intercal(x^t) U^\intercal x^t, g^t \rangle\big]}_{stability} +\text{ other terms } 
    \end{flalign*}
    In this sketch, we will focus solely on bounding the stability error terms. Please refer to Appendix \ref{app:proof-main-bd-mem} for the full proof. Defining the function $\cW:\bR^K \to \bbDelta^N$ as $\cW(G) = \bO_\epsilon\big(-\langle \cY(\cdot)^\intercal U^\intercal \cdot, G \rangle,\bbDelta_N\big)$ allows us to rewrite the stability error term as
    \begin{flalign*}
        \bE\big[\langle g^t, \cW(G^t+\sigma)\rangle \big] - \bE\big[\langle g^t, \cW(G^{t-1}+\sigma)\rangle \big].
    \end{flalign*}
    Note that this expression is a difference of two expectations, which differ from each other solely through a single argument of the averaged function. Regardless of the nature of this function, invoking Proposition \ref{prop:diff_exp_value}  allows us to show that each stability error term scales linearly with $\nu$. While there are $H$ such terms, picking $\nu$ that scales inversely with $\sqrt{H}$ allows us to ensure that the total stability error grows sublinearly in $H$. 
\hfill $\square$

Note that, in the result above, we consider the expected value of the regret because the decisions $\{x^t\}_t$ depend on the random vector $\sigma$ and thus, are random variables themselves. Our result also suggests that it suffices to consider an oracle $\bO_\epsilon$ with $\epsilon \in \cO(T^{-\nicefrac{1}{2}})$ to ensure $\cO(\sqrt{H})$ regret. Crucially, the regret bound we obtained does not depend on the nature of the followers' response function in any way. 
However, it is still important for the leader to possess the ability to precisely evaluate $\cY(\cdot)$ (regardless of its nature),  otherwise she cannot employ the oracle $\bO_\epsilon$ as described in Theorem \ref{thm:main-regret-bd-no-mem}.   

Beyond offering a new perspective on learning in repeated SGs against a sequence of followers, Theorem \ref{thm:main-regret-bd-no-mem} extends existing results by \citet{balcan2015commitment} to settings where followers may not pick best responses. In their approach, they transform the problem \textbf{P1} into an instance of learning from expert advice by observing that it suffices to consider finitely many (albeit, exponential in game parameters) leader strategies when facing best-responding followers. But, this observation no longer holds when facing quantal responding followers, making it unclear how their approach can be extended to this case.     


\subsection{Learning with followers with memory} \label{sec:results-w-memory}

Now, we tackle the main problem we set out to solve and address \textbf{P2} as outlined in Section \ref{sec:followers-w-mem}. In doing so, we first introduce one additional notation and define $\Theta_H = \sum_{t=1}^{H} \frac{1}{b_t} \sum_{\tau=1}^{t} a_{t-\tau} (t-\tau)$. We can state the following 


\begin{thm} \label{thm:main-regret-bd-w-mem}
    Let $\sigma = [\sigma_1,\dots,\sigma_N]$ denote an exponential random vector such that $\sigma_n \stackrel{i.i.d}{\sim} \exp{(\nu)}$ for all $n\in [N]$. 
     Then, setting $\nu= \big(\|U\|_1 (1+L) \sqrt{50 N (\Theta_H+H)}\big)^{-1}$ and picking decisions according to (\ref{eq:approx-FTPL}) with $\cX = \bbDelta_N$, $\rho(\sigma,x) \triangleq - \langle x, \sigma \rangle$ and $f^t = -\langle \cY_{QR}(x)^\intercal U^\intercal x, g^t \rangle$ 
    results in 
    \begin{flalign*}
        \bE[Regret_M(H)] \in \cO\Big(N^{\nicefrac{3}{2}} \|U\|_1 L \sqrt{ \big( H + \Theta_H \big)} + \epsilon H \Big)
    \end{flalign*}
    where the expectation is taken over the distribution of $\sigma$. 
\end{thm}
\textbf{Proof Sketch:} By adding and subtracting some terms and by invoking Theorem \ref{thm:FTPL-general-bound} we can show that 
\begin{flalign*}
    \bE[Regret_M(H)]  = \sum_{t=1}^H & \underbrace{\bE\big[\langle \cY_{QR}^\intercal(x^{t+1}) U^\intercal x^{t+1}, g^t \rangle\big] - \bE\big[\langle \cY^\intercal_{QR}(x^t) U^\intercal x^t, g^t \rangle\big]}_{(A)} \\
    & + \underbrace{\bE\big[\big\langle \cY(x^t)_{QR}^\intercal U^\intercal x^t, g^t \big\rangle\big]-\bE\big[\big\langle \cY(z^t)^\intercal_{QR} U^\intercal x^t, g^t \big\rangle\big]}_{(B)} + \text{ other terms}
\end{flalign*}

 In this sketch, we will focus solely on bounding the stability error terms $(A)$ and $(B)$. Refer Appendix \ref{app:main-regret-bd-w-mem} for the full proof. Term $(A)$ is identical to the stability term in the proof of Theorem \ref{thm:main-regret-bd-no-mem} and as shown there, it can be written as a difference of functions that differ in their argument by an offset. This then allowed us to bound term $(A)$ using Proposition \ref{prop:diff_exp_value}.  However, it is unclear how to extend a similar approach to bound term $(B)$. Prompted by this, we employ a regret analysis methodology in this proof that differs vastly from the one used in proving Theorem \ref{thm:main-regret-bd-no-mem}.  

Restricting the nature of the followers' responses to the quantal response model allows us to employ Corollary \ref{cor:Lip_leader_util} and we can bound the terms $(A)$ and $(B)$ as     
\[(A)\leq L_u \bE[\|x^{t+1}-x^t\|] \text{ and } (B) \leq \frac{1}{b_t} \sum_{\tau=1}^{t-1} a_{t-\tau} L_u' \bE\big[\|x^t-x^\tau\|_1\big]. \] 
Our approach then involves showing that $\bE\big[\|x^t-x^\tau\|_1\big]$ scales linearly with $\nu|t-\tau|$. Consequently, picking a small $\nu$ allows us to ensure that the contribution of each term of the form in $(A)$ and $(B)$ is small. Indeed, picking $\nu$ that scales inversely with $\sqrt{H+\Theta_H}$ allows us to obtain the presented regret bound. \hfill $\square$  

As before, in the result above, we consider the expected value of the regret because the decisions $\{x^t\}_t$ depend on the random vector $\sigma$ and thus, are random variables themselves. Our result suggests that as long as (i) $\Theta_H \in o(H^2)$ and (ii) the oracle $\bO_\epsilon$ is such that $\epsilon \in \cO(H^{-\nicefrac{1}{2}})$,  we can guarantee a sub-linear expected regret.  Also, note that the only property of the QR model we employ in proving this result is the Lipschitz continuity of the response. Thus, our result could be extended to consider other forms of response models that also afford such smoothness properties.

Revisiting our earlier discussion, if we view the decisions $\{x^t\}_t$ as `promises' being made by the leader, our result leads to another interesting observation. After announcing $x^t$, a leader could choose to deviate and play a different strategy. However, Theorem \ref{thm:main-regret-bd-w-mem} suggests that it is advantageous for the leader to honor her promise and play $x^t$ (and earn the expected payoff of \(\big\langle \cY(z^t)^\intercal_{QR} U^\intercal x^t, g^t \big\rangle\)) as it allows her to learn with no-regret.   

Let us now consider two special models of followers' memory and state 
\begin{coro} \label{cor:eff-mem} \textbf{(Bounded Memory)}
    \begin{enumerate}[leftmargin=*]
        \item When the followers have a finite memory of length $B$ and they weigh all past leader decisions equally (i.e. $a_s = 1$ if $s<B$ and $a_s = 0$ otherwise), then $\bE[Regret_M(H)] \in \cO(\sqrt{BH})$. 
        \item When followers weigh past actions by a discount factor ($a_s = \gamma^s$ for some $0<\gamma<1$)$, \bE[Regret_M(H)] \in \cO(\sqrt{H(1-\gamma)^{-1}})$.     
    \end{enumerate}
\end{coro}

Qualitatively, Corollary \ref{cor:eff-mem} argues that for our approach to have sublinear regret we need followers to be `forgetful enough' i.e. the utility of the leader at any stage must not depend strongly on decisions taken far in the past. This is not a surprising result, as it agrees with similar requirements on the memory of an adversary in online learning \cite{arora2012online,hebbar2023online}. 

Note that while $\nu$ depends on $\Theta_H$ in Theorem \ref{thm:main-regret-bd-w-mem}, our approach does not depend on the weights in (\ref{eq:avg_leader-strat}) in any other way. However, if precise knowledge about these weights is not available to the leader, $\Theta_H$ may be difficult to estimate. Fortunately, it suffices to know a suitable upper bound $\bar{\Theta}$ on $\Theta_H$ to make the following  
\begin{coro} \label{robustness}
Picking $\nu= \big(\|U\|_1 (1+L) \sqrt{50 N (\overline\Theta+H)}\big)^{-1}$ in Theorem \ref{thm:main-regret-bd-w-mem} allows us to show that
    \begin{flalign*}
        \bE[Regret_M(H)] \in \cO\Big(N^{\nicefrac{3}{2}} \|U\|_1 L \sqrt{ \big( H + \overline{\Theta} \big)} + \epsilon H \Big)
    \end{flalign*}
\end{coro}

\subsection{Computational aspect of our approaches} \label{sec:comp_aspects}
Let us now address the computational challenges involved in the approaches presented in our work. Note that the design of the oracle employed in Theorems \ref{thm:main-regret-bd-no-mem} and \ref{thm:main-regret-bd-w-mem} strongly depends on the nature of the followers' response functions. Restricting our attention first to the scenario in Theorem \ref{thm:main-regret-bd-no-mem} with best responding followers, we are interested in designing an oracle to approximately solve  
\begin{flalign}
    \max_{x\in \bbDelta_N} \langle \cY_{BR}(x)^\intercal U^\intercal x, \sigma+G^t \rangle = \max_{x\in \bbDelta_N}  \sum_{k\in [K]} (\sigma_k + G^t_k) x^\intercal U y^k_{BR}(x) \label{eq:oracle-BSE}
\end{flalign}
at every time $t$. If we view $\nicefrac{\sigma_k + G^t_k}{\sum_k \sigma_k + G^t_k}$ as the likelihood of the follower being of type $k$, we see that solving (\ref{eq:oracle-BSE}) is equivalent to obtaining the leader's optimal strategy in a Bayesian Stackelberg game with a known distribution over follower types \cite{conitzer2006computing}. This suggests that even obtaining a sufficiently good approximation for (\ref{eq:oracle-BSE}) is NP-hard \cite{letchford2009learning,conitzer2006computing}. However, this equivalence also allows us to use existing MIP solvers \cite{paruchuri2008playing,letchford2009learning} to implement an oracle that solves (\ref{eq:oracle-BSE}). Alternately, we have
\begin{claim} \label{cla:MK_LPs} (Informal)
    An $\epsilon$-optimal solution to (\ref{eq:oracle-BSE}) can be obtained by solving $M^K$ linear programs, and the resulting solution $x^*$ guarantees uniqueness of $\cY_{BR}(x^*)$ without requiring any tie-breaking.  
\end{claim}
The justification of this claim and a discussion comparing it to the computational approach by \citet{balcan2015commitment} can be found in Appendix \ref{app:discuss-MK-LP}. 
Nevertheless, this computational hardness is also encountered in other works that examine learning in Stackelberg games against a sequence of best-responding followers \cite{balcan2015commitment,castiglioni2020online,sessa2020learning}. Much like these studies, our work focuses on developing approaches that allow the leader to quickly learn how to play, rather than on addressing the computational bottlenecks involved.   





On the other hand, when we are faced with memory-less followers that pick quantal response, we wish to approximately solve (\ref{eq:oracle-BSE}) with $\cY_{BR}$ replaced by $\cY_{QR}$. Unfortunately, no polynomial time algorithms are known for computing the optimal leader strategy in a general SG with even a \emph{single} quantal responding follower \cite{cerny2021Dinkelbach,yang2012computing}. Thus, for the purpose of this work, we will employ existing non-convex solvers to implement oracles when faced with quantal responding followers - memory-less or otherwise. Simulation results implementing the methods presented in this section can be found in Appendix \ref{app:sims}.  

\section{Limitations and Future work} \label{sec:limit}

In this work, we sought to design learning algorithms that allow a leader to learn her optimal strategy when faced with followers that deviate from perfect best response. However, our approach assumes the availability of specific optimization oracles which, in practice, are computationally expensive to implement. Further research is needed to develop more computationally efficient oracles, potentially by constraining the class of games under consideration. 

Another limitation of our approach is its reliance on setups with full information feedback, where the type of the follower is revealed at the end of each round. An important direction for future work is to extend our framework to more realistic settings with partial feedback, where the leader only has access to the followers' responses. Addressing this typically involves constructing an unbiased estimator of a follower's type from best responses \cite{balcan2015commitment,castiglioni2020online}. However, it remains to be seen if such an approach extends to the case when followers are quantal responding. 



In this work, we also assumed that the bounded rationality constants are identical across the follower types and is known to the follower. While this assumption simplified our approach in this work, our results are not strongly limited by it. Indeed, we could treat this constant as an unknown (but, parametrized by types) in a follower's payoff model and modify our approach to learn this constant.   

\begin{ack}
This work was supported by the ARO MURI grant W911NF-20-0252
(76582 NSMUR).

\end{ack}


\bibliographystyle{unsrtnat}
\bibliography{neurips_2024}

\begin{thebibliography}{42}
\providecommand{\natexlab}[1]{#1}
\providecommand{\url}[1]{\texttt{#1}}
\expandafter\ifx\csname urlstyle\endcsname\relax
  \providecommand{\doi}[1]{doi: #1}\else
  \providecommand{\doi}{doi: \begingroup \urlstyle{rm}\Url}\fi

\bibitem[Tambe(2011)]{tambe2011security}
Milind Tambe.
\newblock \emph{Security and game theory: algorithms, deployed systems, lessons learned}.
\newblock Cambridge university press, 2011.

\bibitem[Anderson and Engers(1992)]{anderson1992stackelberg}
Simon~P Anderson and Maxim Engers.
\newblock Stackelberg versus cournot oligopoly equilibrium.
\newblock \emph{International Journal of Industrial Organization}, 10\penalty0 (1):\penalty0 127--135, 1992.

\bibitem[Kamenica and Gentzkow(2011)]{kamenica2011bayesian}
Emir Kamenica and Matthew Gentzkow.
\newblock Bayesian persuasion.
\newblock \emph{American Economic Review}, 101\penalty0 (6):\penalty0 2590--2615, 2011.

\bibitem[Hebbar and Langbort(2020)]{hebbar2020stackelberg}
Vijeth Hebbar and Cedric Langbort.
\newblock A stackelberg signaling game for human-uav collaboration in a search-and-rescue context.
\newblock \emph{IFAC-PapersOnLine}, 53\penalty0 (5):\penalty0 297--302, 2020.

\bibitem[Massicot and Langbort(2019)]{massicot2019public}
Olivier Massicot and Cedric Langbort.
\newblock Public signals and persuasion for road network congestion games under vagaries.
\newblock \emph{IFAC-PapersOnLine}, 51\penalty0 (34):\penalty0 124--130, 2019.

\bibitem[Farokhi et~al.(2016)Farokhi, Teixeira, and Langbort]{farokhi2016estimation}
Farhad Farokhi, Andr{\'e}~MH Teixeira, and C{\'e}dric Langbort.
\newblock Estimation with strategic sensors.
\newblock \emph{IEEE Transactions on Automatic Control}, 62\penalty0 (2):\penalty0 724--739, 2016.

\bibitem[Zrnic et~al.(2021)Zrnic, Mazumdar, Sastry, and Jordan]{zrnic2021leads}
Tijana Zrnic, Eric Mazumdar, Shankar Sastry, and Michael Jordan.
\newblock Who leads and who follows in strategic classification?
\newblock \emph{Advances in Neural Information Processing Systems}, 34:\penalty0 15257--15269, 2021.

\bibitem[Letchford et~al.(2009)Letchford, Conitzer, and Munagala]{letchford2009learning}
Joshua Letchford, Vincent Conitzer, and Kamesh Munagala.
\newblock Learning and approximating the optimal strategy to commit to.
\newblock In \emph{Algorithmic Game Theory: Second International Symposium, SAGT 2009, Paphos, Cyprus, October 18-20, 2009. Proceedings 2}, pages 250--262. Springer, 2009.

\bibitem[Blum et~al.(2014)Blum, Haghtalab, and Procaccia]{blum2014learning}
Avrim Blum, Nika Haghtalab, and Ariel~D Procaccia.
\newblock Learning optimal commitment to overcome insecurity.
\newblock \emph{Advances in Neural Information Processing Systems}, 27, 2014.

\bibitem[Balcan et~al.(2015)Balcan, Blum, Haghtalab, and Procaccia]{balcan2015commitment}
Maria-Florina Balcan, Avrim Blum, Nika Haghtalab, and Ariel~D Procaccia.
\newblock Commitment without regrets: Online learning in stackelberg security games.
\newblock In \emph{Proceedings of the sixteenth ACM conference on economics and computation}, pages 61--78, 2015.

\bibitem[Sessa et~al.(2020)Sessa, Bogunovic, Kamgarpour, and Krause]{sessa2020learning}
Pier~Giuseppe Sessa, Ilija Bogunovic, Maryam Kamgarpour, and Andreas Krause.
\newblock Learning to play sequential games versus unknown opponents.
\newblock \emph{Advances in neural information processing systems}, 33:\penalty0 8971--8981, 2020.

\bibitem[Castiglioni et~al.(2020)Castiglioni, Celli, Marchesi, and Gatti]{castiglioni2020online}
Matteo Castiglioni, Andrea Celli, Alberto Marchesi, and Nicola Gatti.
\newblock Online bayesian persuasion.
\newblock \emph{Advances in Neural Information Processing Systems}, 33:\penalty0 16188--16198, 2020.

\bibitem[Marecki et~al.(2012)Marecki, Tesauro, and Segal]{marecki2012playing}
Janusz Marecki, Gerry Tesauro, and Richard Segal.
\newblock Playing repeated stackelberg games with unknown opponents.
\newblock In \emph{Proceedings of the 11th International Conference on Autonomous Agents and Multiagent Systems-Volume 2}, pages 821--828, 2012.

\bibitem[Blum and Kalai(1997)]{blum1997universal}
Avrim Blum and Adam Kalai.
\newblock Universal portfolios with and without transaction costs.
\newblock In \emph{Proceedings of the Tenth Annual Conference on Computational Learning Theory}, pages 309--313, 1997.

\bibitem[Velicheti et~al.(2024)Velicheti, Bastopcu, Etesami, and Ba{\c{s}}ar]{velicheti2024learning}
Raj~Kiriti Velicheti, Melih Bastopcu, S~Rasoul Etesami, and Tamer Ba{\c{s}}ar.
\newblock Learning how to strategically disclose information.
\newblock \emph{arXiv preprint arXiv:2403.08741}, 2024.

\bibitem[Xu et~al.(2016)Xu, Tran-Thanh, and Jennings]{xu2016playing}
Haifeng Xu, Long Tran-Thanh, and Nick Jennings.
\newblock Playing repeated security games with no prior knowledge.
\newblock In \emph{AAMAS'16: Proceedings of the 2016 International Conference on Autonomous Agents \& Multiagent Systems}, pages 104--112. ACM Press, 2016.

\bibitem[Lin and Chen(2024)]{lin2024persuading}
Tao Lin and Yiling Chen.
\newblock Persuading a learning agent.
\newblock \emph{arXiv preprint arXiv:2402.09721}, 2024.

\bibitem[Deng et~al.(2019)Deng, Schneider, and Sivan]{deng2019strategizing}
Yuan Deng, Jon Schneider, and Balasubramanian Sivan.
\newblock Strategizing against no-regret learners.
\newblock \emph{Advances in neural information processing systems}, 32, 2019.

\bibitem[Cesa-Bianchi et~al.(2013)Cesa-Bianchi, Dekel, and Shamir]{cesa2013online}
Nicolo Cesa-Bianchi, Ofer Dekel, and Ohad Shamir.
\newblock Online learning with switching costs and other adaptive adversaries.
\newblock \emph{Advances in Neural Information Processing Systems}, 26, 2013.

\bibitem[Fiez et~al.(2020)Fiez, Chasnov, and Ratliff]{fiez2020implicit}
Tanner Fiez, Benjamin Chasnov, and Lillian Ratliff.
\newblock Implicit learning dynamics in stackelberg games: Equilibria characterization, convergence analysis, and empirical study.
\newblock In \emph{International Conference on Machine Learning}, pages 3133--3144. PMLR, 2020.

\bibitem[Haghtalab et~al.(2024)Haghtalab, Podimata, and Yang]{haghtalab2024calibrated}
Nika Haghtalab, Chara Podimata, and Kunhe Yang.
\newblock Calibrated stackelberg games: Learning optimal commitments against calibrated agents.
\newblock \emph{Advances in Neural Information Processing Systems}, 36, 2024.

\bibitem[Haghtalab et~al.(2022)Haghtalab, Lykouris, Nietert, and Wei]{haghtalab2022learning}
Nika Haghtalab, Thodoris Lykouris, Sloan Nietert, and Alexander Wei.
\newblock Learning in stackelberg games with non-myopic agents.
\newblock In \emph{Proceedings of the 23rd ACM Conference on Economics and Computation}, pages 917--918, 2022.

\bibitem[McKelvey and Palfrey(1995)]{mckelvey1995quantal}
Richard~D McKelvey and Thomas~R Palfrey.
\newblock Quantal response equilibria for normal form games.
\newblock \emph{Games and economic behavior}, 10\penalty0 (1):\penalty0 6--38, 1995.

\bibitem[Feng et~al.(2024)Feng, Ho, and Tang]{feng2024rationality}
Yiding Feng, Chien-Ju Ho, and Wei Tang.
\newblock Rationality-robust information design: Bayesian persuasion under quantal response.
\newblock In \emph{Proceedings of the 2024 Annual ACM-SIAM Symposium on Discrete Algorithms (SODA)}, pages 501--546. SIAM, 2024.

\bibitem[Yang et~al.(2012)Yang, Ordonez, and Tambe]{yang2012computing}
Rong Yang, Fernando Ordonez, and Milind Tambe.
\newblock Computing optimal strategy against quantal response in security games.
\newblock In \emph{AAMAS}, pages 847--854. Citeseer, 2012.

\bibitem[Sutton and Barto(2018)]{sutton2018reinforcement}
Richard~S Sutton and Andrew~G Barto.
\newblock \emph{Reinforcement learning: An introduction}.
\newblock MIT press, 2018.

\bibitem[Goodfellow et~al.(2016)Goodfellow, Bengio, and Courville]{goodfellow2016deep}
Ian Goodfellow, Yoshua Bengio, and Aaron Courville.
\newblock \emph{Deep learning}.
\newblock MIT press, 2016.

\bibitem[Arora et~al.(2012{\natexlab{a}})Arora, Hazan, and Kale]{arora2012multiplicative}
Sanjeev Arora, Elad Hazan, and Satyen Kale.
\newblock The multiplicative weights update method: a meta-algorithm and applications.
\newblock \emph{Theory of computing}, 8\penalty0 (1):\penalty0 121--164, 2012{\natexlab{a}}.

\bibitem[Wu et~al.(2022)Wu, Shen, Fang, and Xu]{wu2022inverse}
Jibang Wu, Weiran Shen, Fei Fang, and Haifeng Xu.
\newblock Inverse game theory for stackelberg games: the blessing of bounded rationality.
\newblock \emph{Advances in Neural Information Processing Systems}, 35:\penalty0 32186--32198, 2022.

\bibitem[Arora et~al.(2012{\natexlab{b}})Arora, Dekel, and Tewari]{arora2012online}
Raman Arora, Ofer Dekel, and Ambuj Tewari.
\newblock Online bandit learning against an adaptive adversary: from regret to policy regret.
\newblock \emph{arXiv preprint arXiv:1206.6400}, 2012{\natexlab{b}}.

\bibitem[Anava et~al.(2015)Anava, Hazan, and Mannor]{anava2015online}
Oren Anava, Elad Hazan, and Shie Mannor.
\newblock Online learning for adversaries with memory: price of past mistakes.
\newblock \emph{Advances in Neural Information Processing Systems}, 28, 2015.

\bibitem[Hebbar and Langbort(2023)]{hebbar2023online}
Vijeth Hebbar and Cedric Langbort.
\newblock Online decision making with history-average dependent costs (extended).
\newblock \emph{arXiv preprint arXiv:2312.06641}, 2023.

\bibitem[Agarwal et~al.(2019)Agarwal, Gonen, and Hazan]{agarwal2019learning}
Naman Agarwal, Alon Gonen, and Elad Hazan.
\newblock Learning in non-convex games with an optimization oracle.
\newblock In \emph{Conference on Learning Theory}, pages 18--29. PMLR, 2019.

\bibitem[Suggala and Netrapalli(2020)]{suggala2020online}
Arun~Sai Suggala and Praneeth Netrapalli.
\newblock Online non-convex learning: Following the perturbed leader is optimal.
\newblock In \emph{Algorithmic Learning Theory}, pages 845--861. PMLR, 2020.

\bibitem[Kalai and Vempala(2005)]{kalai2005efficient}
Adam Kalai and Santosh Vempala.
\newblock Efficient algorithms for online decision problems.
\newblock \emph{Journal of Computer and System Sciences}, 71\penalty0 (3):\penalty0 291--307, 2005.

\bibitem[Hazan(2022)]{hazan2016introduction}
Elad Hazan.
\newblock \emph{Introduction to online convex optimization}, chapter~5.
\newblock The MIT Press, 2 edition, 2022.

\bibitem[Cesa-Bianchi and Lugosi(2006)]{cesa2006prediction}
Nicolo Cesa-Bianchi and G{\'a}bor Lugosi.
\newblock \emph{Prediction, learning, and games}.
\newblock Cambridge university press, 2006.

\bibitem[Fudenberg and Levine(1998)]{fudenberg1998theory}
Drew Fudenberg and David~K Levine.
\newblock \emph{The theory of learning in games}, volume~2.
\newblock MIT press, 1998.

\bibitem[Mertikopoulos and Sandholm(2016)]{mertikopoulos2016learning}
Panayotis Mertikopoulos and William~H Sandholm.
\newblock Learning in games via reinforcement and regularization.
\newblock \emph{Mathematics of Operations Research}, 41\penalty0 (4):\penalty0 1297--1324, 2016.

\bibitem[Conitzer and Sandholm(2006)]{conitzer2006computing}
Vincent Conitzer and Tuomas Sandholm.
\newblock Computing the optimal strategy to commit to.
\newblock In \emph{Proceedings of the 7th ACM conference on Electronic commerce}, pages 82--90, 2006.

\bibitem[Paruchuri et~al.(2008)Paruchuri, Pearce, Marecki, Tambe, Ordonez, and Kraus]{paruchuri2008playing}
Praveen Paruchuri, Jonathan~P Pearce, Janusz Marecki, Milind Tambe, Fernando Ordonez, and Sarit Kraus.
\newblock Playing games for security: An efficient exact algorithm for solving bayesian stackelberg games.
\newblock In \emph{Proceedings of the 7th international joint conference on Autonomous agents and multiagent systems-Volume 2}, pages 895--902, 2008.

\bibitem[\v{C}ern\'{y} et~al.(2021)\v{C}ern\'{y}, Lis\'{y}, Bo\v{s}ansk\'{y}, and An]{cerny2021Dinkelbach}
Jakub \v{C}ern\'{y}, Viliam Lis\'{y}, Branislav Bo\v{s}ansk\'{y}, and Bo~An.
\newblock Dinkelbach-type algorithm for computing quantal stackelberg equilibrium.
\newblock In \emph{Proceedings of the Twenty-Ninth International Joint Conference on Artificial Intelligence}, IJCAI'20, 2021.
\newblock ISBN 9780999241165.

\end{thebibliography}

\medskip






\newpage

\section{Proofs for Various Results} \label{app:proofs}

\subsection{Proof of Proposition \ref{prop:diff_exp_value}}
    Let $\cU=[0,\frac{2}{\nu}]^n$ denote the support of $\sigma$ and let $\cU_c=[c,c+\frac{2}{\nu}]^n$ denote the support of the shifted random variable $\sigma'=c+\sigma$. Then we have  
    \begin{flalign*}
        \big| \bE[f(G+c+\sigma) - f(G+\sigma)] \big| = & \bigg|\int_{\cU_c} \frac{1}{Vol(\cU_c)} f(G+z) dz - \int_\cU \frac{1}{Vol(\cU)} f(G+z) dz    \bigg|  \\
        = & \frac{1}{Vol(\cU)} \bigg| \int_{\cU_c\setminus \cU} f(G+z) dz  - \int_{\cU\setminus \cU_c} f(G+z)dz \bigg| \\
        \stackrel{(a)}{\leq} &  \frac{1}{Vol(\cU)} \bigg(\int_{\cU_c\setminus \cU} B dz  + \int_{\cU\setminus \cU_c} B dz \bigg) \\
        \stackrel{(b)}{=} & 2\bigg(\frac{\nu}{2}\bigg)^n B \; \big(Vol(\cU/\cU_c)\big) \stackrel{(c)}{\leq} 2\bigg(\frac{\nu}{2}\bigg)^n B \sum_i |c_i| \bigg(\frac{2}{\nu}\bigg)^{n-1} \leq \nu B \|c\|_1 
    \end{flalign*}
    where the inequality in $(a)$ results from triangle inequality, the fact that $|\int f| \leq \int |f|$ and from boundedness of $B$. The equality in $(b)$ is by exploiting the symmetry of $\cU/\cU_c$ and $\cU_c/\cU$. The inequality in $(c)$ was obtained by counting the volume between the surface of $\cU_c$ that is cutting through $\cU$ and the corresponding parallel surface of $\cU$. Note that in summing up these volumes some regions are double counted and thus the inequality.  
\subsection{Proof for Claim \ref{claim:diff_y_bd}}
We begin by analyzing the gradient of the quantal response function. Dropping the sub-script `QR' and super-script $k$ for brevity and by focusing on the $i$ element of $y$ we have
\begin{flalign*}
    \nabla y_i(x) = & \frac{\eta V_i \exp{\big(\eta   x^\intercal {V_i}\big)}}{\sum_j \exp{\big(\eta  x^\intercal {V_j}\big)}} - \frac{\exp{\big(\eta   x^\intercal {V_i}\big)}}{\sum_j \exp{\big(\eta  x^\intercal {V_j}\big)}} \bigg(\frac{\sum_j \eta V_j \exp{\big(\eta   x^\intercal {V_j}\big)}}{\sum_j \exp{\big(\eta  x^\intercal {V_j}\big)}}\bigg) \\
    = &\eta V_i y_i(x) - \eta y_i(x) \sum_j V_j y_j(x) = \eta y_i(x) \big( V_i - V y(x)\big). 
\end{flalign*}
We can bound the 1-norm of this gradient simply as
\begin{flalign*}
    \|\nabla y_i(x)\|_1 = \eta y_i(x) \big\|V \big(e_i - y(x)\big)\big\|_1 \leq \eta y_i(x) \|V\|_1 \|e_i - y(x)\|_1 \leq 2\eta y_i(x) \|V\|_1    
\end{flalign*}
where $e_i$ denotes the $i^{th}$ standard basis vector in $\bR^M$. Now to prove our claim consider 
\begin{flalign*}
    \|y(x) - y(z)\|_1 = \sum_{i} |y_{i}(x) - y_i(z)| \stackrel{(a)}{\leq} 
    \sum_i 2\eta y_i(x) \|V\|_1 \|x-z\|_\infty = 2\eta \|V\|_1 \|x-z\|_\infty  
\end{flalign*}
where $(a)$ results from a combination of applying mean value theorem on the differentiable function $y_i$, Holder's inequality and using the bound obtained above. 
\hfill $\square$

\subsection{Proof of Corollary \ref{cor:Lip_leader_util}}
To avoid cluttered notation, in this proof, we will drop the $QR$ subscript and assume that $\cY$ captures the quantal response of followers. 
\begin{flalign*}
    |x^\intercal U \cY(x) g - z^\intercal U \cY(z) g| & = |(x - z)^\intercal U \cY(x) g + \overbrace{z^\intercal U (\cY(x) - \cY(z)) g}^{(A)}| \\
    & \stackrel{(a)}{\leq}  \sum_{k=1}^K g_k |(x - z)^\intercal U y^k(x)| + \sum_{k=1}^K g_k \big|z^\intercal U \big(y^k(x) - y^k(z)\big)\big| \\   
    & \stackrel{(b)}{\leq} \sum_{k=1}^K g_k \|(x - z)\|_\infty \|U y^k(x)\|_1 + \sum_{k=1}^K g_k \|z\|_\infty \big\|U \big(y^k(x) - y^k(z)\big)\big\|_1 \\
    & \stackrel{}{\leq} \|x - z\|_\infty \sum_{k=1}^K g_k \| U \|_1 \|y^k(x)\|_1 + \|z\|_\infty \sum_{k=1}^K g_k \|U\|_1 \|y^k(x)- y^k(x)\|_1
\end{flalign*}
Inequality $(a)$ above results from noting that $\cY(x) g = \sum_{k=1}^K g_k y^k(x)$ and from subsequently applying triangle inequality.  Inequality $(b)$ results from the application of Holder's inequality. Invoking Claim \ref{claim:diff_y_bd} and using $\|\cdot\|_\infty \leq \|\cdot\|_1$ over finite dimensional vectors allows us to further bound 
\begin{flalign*}
    |x^\intercal U \cY(x) g - z^\intercal U \cY(z) g| & = \|x - z\|_\infty \sum_{k=1}^K g_k \| U \|_1 + \|z\|_1 \sum_{k=1}^K g_k \|U\|_1 L \|x-z\|_\infty \\
    & \stackrel{}{\leq}  (1+L) \| U \|_1 \bigg(\sum_k g_k\bigg)  \|x - z\|_\infty  
\end{flalign*}
and the result follows. The bound in (\ref{eq:aux-leader-util-bd}) can be obtained similarly by considering only the term (A) and tracking its bounds throughout.  

\subsection{Proof of Theorem \ref{thm:FTPL-general-bound}}
Before we prove Theorem \ref{thm:FTPL-general-bound} with a non-zero regularizer function, let us first assume that $\rho = 0$. For for decisions made according to 
\begin{flalign} \label{eq:FTL-oracle}
    \forall t > 1, x^{t} = \bO_{\epsilon}\bigg(\sum_{\tau=1}^{t-1} f^\tau \bigg) \quad  
    S.T \; \sum_{\tau=1}^{t-1} f^\tau(x^{t}) \leq \min_{x\in \cX} \sum_{\tau=1}^{t-1} f^t(x) + \epsilon
\end{flalign}
we can state and prove the following 
\begin{lem} \label{lem:FTL-regret-bd}
    When $\{x^t\}_{t>1}$ is generated according to (\ref{eq:FTL-oracle}) we can show that
    \[
        \sum_{t=1}^H f^t(x^{t+1}) \leq \min_{x\in \cX} \sum_{t=1}^H f^t(x) + \epsilon H 
    \]       
\end{lem}
\textbf{Proof:} To show this we will employ induction. Clearly, from (\ref{eq:FTL-oracle}) we have $f^1(x^{2}) \leq \min_{x\in \cX} f^1(x)  + \epsilon.$ Now, let us suppose that for some $1\leq h<H$ we have 
\[\sum_{t=1}^h  f^t(x^{t+1}) \leq \min_{x\in \cX} \sum_{t=1}^h f^t(x)  + \epsilon h.\] 
Then we have 
\begin{flalign*}
    \sum_{t=1}^{h+1}  f^t(x^{t+1}) & = \sum_{t=1}^{h}  f^t(x^{t+1}) + f^{h+1}(x^{h+2}) \stackrel{(a)}{\leq}  \min_{x\in \cX} \sum_{t=1}^h f^t(x) + f^{h+1}(x^{h+2}) + \epsilon h \\
    & \leq \sum_{t=1}^h f^t(x^{h+2}) + f^{h+1}(x^{h+2}) + \epsilon h \stackrel{(b)}{\leq} \min_{x\in \cX}\sum_{t=1}^{h+1} f^t(x) + \epsilon + \epsilon h
\end{flalign*}
where $(a)$ results from the inductive hypothesis and $(b)$ follows from the definition of the oracle in (\ref{eq:FTL-oracle}). 
\hfill $\square$

We can now prove Theorem \ref{thm:FTPL-general-bound}. To do so, let us define $\tilde{f}^0(\cdot) = R(\sigma,\cdot)$ and $\tilde{f}^t(\cdot) = {f}^t(\cdot)$. Then, we can make the observation that picking ${x^t}$ according to (\ref{eq:approx-FTPL}) is equivalent to picking decision according (\ref{eq:FTL-oracle}) with cost functions as $\{\tilde{f}^t\}_{t\geq 0}$ with time starting from $0$ instead of $1$. Consequently, by monotonicity of expectation, for any $x'\in \cX$ we have
\begin{flalign*}
    \bE\bigg[\rho(\sigma,x')+\sum_{t=1}^{H}  f^t(x')\bigg] & = \bE\bigg[\sum_{t=0}^{H}  \tilde f^t(x')\bigg] \geq \bE\bigg[\min_{x\in \cX} \sum_{t=0}^{H}  \tilde f^t(x)\bigg] \\
    & \stackrel{(c)}{\geq} \bE[\rho(\sigma,x^1)] + \sum_{t=1}^H \bE[f^t(x^{t+1})] - \epsilon (H+1)
\end{flalign*}
where the expectations above are taken over the distribution of $\sigma$ and
$(c)$ results from Lemma \ref{lem:FTL-regret-bd}. Rearranging terms, subtracting $\sum_{t=1}^H \bE[f^t(x^t)]$ throughout the inequation and picking $x'$ as the best-in-hindsight strategy completes the proof. \hfill $\square$

\subsection{Proof of Theorem \ref{thm:main-regret-bd-no-mem}} \label{app:proof-main-bd-mem}
First, we restate the theorem with some additional details added in and then provide the complete proof for it.
\begin{thm*} 
    Let $\sigma = [\sigma_1,\dots,\sigma_K]$ denote a uniform random vector such that $\sigma_k \stackrel{i.i.d}{\sim} U[0,\nicefrac{2}{\nu}]$ for all $k\in [K]$. Let $\rho(\sigma,x) \triangleq - \langle \cY(x)^\intercal U^\intercal x, \sigma\rangle$.
    Then, setting $\nu = \sqrt{\nicefrac{K}{H}}$ and picking decisions according to 
    \begin{equation}
        \forall t>1, x^{t} = \bO_{\epsilon}\Big(-\langle \cY(\cdot)^\intercal U^\intercal \cdot, G^{t-1} \rangle + \rho(\sigma,\cdot),\bbDelta_N\Big) \text{ and } x^1 = \bO_{\epsilon}\big( \rho(\sigma,\cdot),\bbDelta_N\big) \label{eq:FTPL-no-memory}
    \end{equation}
    allows us to bound the expected value of regret in (\ref{eq:regret-memory}) as 
    \begin{flalign*}
        \bE[Regret(H)] \leq 2 \overline{U} \sqrt{KH} + \epsilon (H+1)
    \end{flalign*}
    where the expectation is taken over the distribution of $\sigma$.
\end{thm*}

\textbf{Proof:} Note that (\ref{eq:FTPL-memory}) is equivalent to (\ref{eq:approx-FTPL}) when $f^t = -\langle \cY(x)^\intercal U^\intercal x, g^t \rangle$ and $\cX = \bbDelta^N$. Then, invoking Theorem \ref{thm:FTPL-general-bound} gives us 
\begin{flalign*}
         \bE\big[Regret(H)\big] \leq & \sum_{t=1}^H \underbrace{\bE\big[\langle \cY^\intercal(x^{t+1}) U^\intercal x^{t+1}, g^t \rangle\big] - \bE\big[\langle \cY^\intercal(x^t) U^\intercal x^t, g^t \rangle\big]}_{(A)} \\
        & + \underbrace{\bE\big[\langle \cY^\intercal(x^1) U^\intercal x^1 - \cY^\intercal(x') U^\intercal x' ,\sigma \rangle\big]}_{(B)} + \epsilon (H+1).
    \end{flalign*}
    Focusing first on term $(B)$, a simple application of the Holder's inequality gives us.
    \begin{flalign*}
        (B) & \leq \bE\big[\| {x^1}^\intercal U \cY(x^1) - {x'}^\intercal U \cY(x') \|_\infty \| \sigma \|_1 \big] \stackrel{(a)}{\leq} \overline{U} \bE[\|\sigma\|_1] = \overline{U} \frac{K}{\epsilon}
    \end{flalign*}
    where $(a)$ follows from noting that each column of $\cY$ is an element of $\bbDelta_M$ and thus, ${x}^\intercal U \cY(x)$ is a non-negative vector with each element being atmost $\overline{U}$. 
    
    Shifting our focus now to terms of the form in $(A)$, let us first define the function $\cW:\bR^K \to \bbDelta^N$ as 
    \[\cW(G) = \bO_\epsilon\big(-\langle \cY(\cdot)^\intercal U^\intercal \cdot, G \rangle\big)\]
    If we define $G^0 = [0,\dots,0]^\intercal \in \bR^K$, then at any time $t\geq 1$, we observe that $x^t=\cW(G^{t-1}+\sigma)$. Then term (A) can be re-written as 
    \begin{flalign*}
        \bE\big[\langle \cY^\intercal(x^{t+1}) U^\intercal x^{t+1}- \langle \cY^\intercal(x^t) U^\intercal x^t, g^t \rangle\big] & = \bE\big[\langle g^t, \cW(G^t+\sigma)\rangle \big] - \bE\big[\langle g^t, \cW(G^{t-1}+\sigma)\rangle \big].
    \end{flalign*}
    Invoking Proposition \ref{prop:diff_exp_value} then allows us to upperbound term $(A)$ as 
    \begin{flalign*}
        & \bE\big[\langle g^t, \cW(G^t+\sigma)\rangle \big] - \bE\big[\langle g^t, \cW(G^{t-1}+\sigma)\rangle \big] \\
        & \leq \bigg| \bE\big[\langle g^t, \cW(G^{t-1}+g^t+\sigma)\rangle \big] - \bE\big[\langle g^t, \cW(G^{t-1}+\sigma)\rangle \big] \bigg| \\
        &\leq \nu \overline{U} \|g_t\|_1 = \nu\overline{U}.
    \end{flalign*}
    Since there are $H$ terms of the form (A) we get the regret bound as
    \[\bE[Regret(H)] \leq \nu H \overline{U} + \overline{U} \frac{K}{\nu} + \epsilon (H+1).\]
    Setting $\nu = \sqrt{\frac{K}{H}}$ in this bound completes the proof. \hfill $\square$

\subsection{Proof of Corollary \ref{cor:eff-mem}}
\begin{enumerate}
    \item We merely need to compute  
    \begin{flalign*}
        \Theta_H & = \sum_{t=1}^{H} \frac{1}{b_t} \sum_{\tau=1}^{t} a^{t-\tau} (t-\tau) = \frac{(B-1)(H-B+2)}{4}
    \end{flalign*}
    and the result follows. 
    \item Once again we compute  
    \begin{flalign*}
        \Theta_H & = \sum_{t=1}^{H} \frac{1}{b_t} \sum_{\tau=1}^{t} a^{t-\tau} (t-\tau) = \sum_{t=1}^H \frac{\gamma}{1-\gamma}\bigg(\frac{t\gamma^{t-1}(1-\gamma)}{(1-\gamma^t)} + 1\bigg) \\
        & \leq \sum_{t=1}^H \frac{\gamma}{1-\gamma}\big(t \gamma^{t-1} + 1\big) = \frac{\gamma}{1-\gamma} \bigg(H + \frac{(H+1)\gamma^H (1-\gamma)+ 1- \gamma^{H+1}}{(1-\gamma)^2}\bigg)
    \end{flalign*}
    and the result follows from observing that $\Theta_H+H = \frac{H}{1-\gamma} + \cO\bigg(\frac{H\gamma^{H+1}}{(1-\gamma)^3}\bigg)$.
\end{enumerate}
\section{Proof of Theorem \ref{thm:main-regret-bd-w-mem}} \label{app:main-regret-bd-w-mem} 
We begin by first proving some useful results that will aid us with our regret analysis.

\subsection{Preliminaries for Theorem \ref{thm:main-regret-bd-w-mem}}

\begin{lem} \label{lem:generalized-stability-bd}
    Let $\{f^t\}_{t=1}^H$ be a sequence of L-Lipshitz  continuous (w.r.t $L_1$-norm) cost functions that map a convex set $\cX \subset\bR^N  \to \bR$ and let $\sigma = [\sigma_1,\dots,\sigma_N]$ denote an exponential random vector such that $\sigma_n \stackrel{i.i.d}{\sim} \exp{(\nu)}$ for all $n\in [N]$. Let $D$ be the $l_\infty$ diameter of the \emph{convex} 
    domain $\cX$. Let $\{x_t\}_{t=1}^H$ be generated according to 
    \begin{flalign} 
        \forall t > 1, \; x^{t} = \bO_{\epsilon}\bigg(\sum_{\tau=1}^{t-1} f^\tau - \langle \sigma,\cdot\rangle \bigg) \quad  
        \text{ and } x^1 = \bO_{\epsilon}\big( - \langle \sigma,\cdot\rangle \big) \label{eq:non-cvx-learn-opt-oracle}
    \end{flalign}
    then for any $h>0$, the expected value of the stability error terms can be bounded above as 
    \begin{flalign} 
        \bE[\|x_{t+h}(\sigma)-x_{t}(\sigma)\|_1]  \leq    50\nu hLN^2D + 6 \frac{\epsilon}{hL} \label{eq:stab_bound} 
    \end{flalign}
    where the expectation is taken over the distribution of $\sigma$.
\end{lem}

    

This proof will closely parallel the approach taken by \citet{suggala2020online}(Theorem 1), where they shows a similar result for the special case when $h=1$. Before we dive into the proof of Lemma \ref{lem:generalized-stability-bd}, we present two more helper results.

\begin{lem}[\citet{suggala2020online}]\label{lem:monotonicity-homogenous}
    Let $x_t(\sigma)$ denote the decisions generated according to (\ref{eq:non-cvx-learn-opt-oracle}), highlighting their dependence on the perturbation $\sigma$. Let $x_{t,i}$ denote the $i^{th}$ component of $x_t$ and let $e_i$ denote the $i^{th}$ standard basis vector in $\bR^N$. Then for any $c>0$
    \[x_{t,i}(\sigma+ce_i) \geq x_{t,i}(\sigma) - 2\frac{\epsilon}{c}.\]
\end{lem}

\textbf{Proof:} Let $F^t = \sum_{\tau<t} f^\tau$ and let $\sigma'=\sigma + ce_i$. 
Then we have
\begin{flalign*}
    F^{t}(x_t(\sigma)) - \langle \sigma,x_t(\sigma) \rangle &  \stackrel{(a)}{\leq} 
    F^{t}(x_t(\sigma')) - \langle \sigma,x_t(\sigma') \rangle + \epsilon \\ 
    & \stackrel{(b)}{=} F^{t}(x_t(\sigma')) - \langle \sigma',x_t(\sigma') \rangle + cx_{t,i}(\sigma')+ \epsilon \\ 
    & \stackrel{(c)}{\leq} F^{t}(x_t(\sigma)) - \langle \sigma',x_t(\sigma) \rangle + cx_{t,i}(\sigma') + 2\epsilon \\ 
    & \stackrel{(d)}{=} F^{t}(x_t(\sigma)) - \langle \sigma,x_t(\sigma) \rangle + c\big(x_{t,i}(\sigma') - x_{t,i}(\sigma)\big)+2\epsilon 
\end{flalign*}
where inequality $(a)$ and $(c)$ result from the definition of the oracle $\bO_{\epsilon}$ and equalities $(b)$ and $(d)$ result simple from the relation that $\sigma'=\sigma+ce_i$. The result follows from comparing the RHS of $(d)$ and the LHS of $(a)$.
\hfill $\square$

All $\|\cdot\|$ below are $L1$-norm unless otherwise specified. 
\begin{lem}\label{lem:monotonicity-heterogenous}
    Let $x_t(\sigma)$ denote the decisions generated according to (\ref{eq:non-cvx-learn-opt-oracle}), highlighting their dependence on the perturbation $\sigma$. Let $x_{t,i}$ denote the $i^{th}$ component of $x_t$ and let $e_i$ denote the $i^{th}$ standard basis vector in $\bR^N$. Now for any $p>0$ and $q>0$ if
    \begin{flalign}
        \|x_t(\sigma)-x_{t+h}(\sigma)\| \leq pN|x_{t,i}(\sigma)-x_{t+h,i}(\sigma)| \label{eq:norm_bd_assum}
    \end{flalign} 
    then for $\sigma'=\sigma+qLN e_i$
    \[\min\{x_{t,i}(\sigma'),x_{t+h,i}(\sigma')\} \geq \max\{x_{t,i}(\sigma),x_{t+h,i}(\sigma)\} -\frac{ph}{q}|x_{t,i}(\sigma)-x_{t+h,i}(\sigma)| - \frac{3\epsilon}{q LN}\]
\end{lem}
\textbf{Proof:} Let $F^t = \sum_{\tau< t} f^t$, $f^{t:t+h} = \sum_{\tau=t}^{t+h-1} f^\tau$ and $\sigma'=\sigma + qLN e_i$. 
Then we have  
\begin{flalign*}
     & F^t(x_t(\sigma))-\langle \sigma,x_t(\sigma) \rangle + f^{t:t+h}(x_t(\sigma)) \\
    \stackrel{(a)}{\leq} & 
    F^t(x_{t+h}(\sigma))-\langle \sigma,x_{t+h}(\sigma) \rangle + f^{t:t+h}(x_t(\sigma)) + \epsilon \\
    \stackrel{(b)}{\leq} & 
    F^t(x_{t+h}(\sigma))-\langle \sigma,x_{t+h}(\sigma) \rangle + f^{t:t+h}(x_{t+h}(\sigma)) + Lh\|x_t(\sigma) - x_{t+h}(\sigma)\|+ \epsilon \\
    \stackrel{(c)}{\leq} & 
    F^t(x_{t+h}(\sigma))-\langle \sigma,x_{t+h}(\sigma) \rangle + f^{t:t+h}(x_{t+h}(\sigma)) + hpNL|x_{t,i}(\sigma)-x_{t+h,i}(\sigma)|+ \epsilon 
\end{flalign*}
here inequality $(a)$ result from the definition of the oracle $\bO_{\epsilon}$, $(b)$ results from the Lipschitz property of $f^\tau$ and $(c)$ results from (\ref{eq:norm_bd_assum}). On the other hand, we have 
\begin{flalign*}
     & F^t(x_t(\sigma))-\langle \sigma,x_t(\sigma) \rangle + f^{t:t+h}(x_t(\sigma)) \\
    \stackrel{(d)}{=} &  F^t(x_t(\sigma))-\langle \sigma',x_t(\sigma) \rangle + f^{t:t+h}(x^t(\sigma)) + qLN x_{t,i}(\sigma) \\
    \stackrel{(e)}{\geq} & F^{t+h}(x_{t+h}(\sigma'))-\langle \sigma',x_{t+h}(\sigma') \rangle + qLN x_{t,i}(\sigma) - \epsilon  \\
    \stackrel{(f)}{=} & F^{t+h}(x_{t+h}(\sigma'))-\langle \sigma,x_{t+h}(\sigma') \rangle + qLN \big(x_{t,i}(\sigma) - x_{t+h,i}(\sigma')\big) - \epsilon\\
    \stackrel{(g)}{\geq} & F^{t+h}(x_{t+h}(\sigma))-\langle \sigma,x_{t+h}(\sigma) \rangle + qLN \big(x_{t,i}(\sigma) - x_{t+h,i}(\sigma')\big) - 2\epsilon
\end{flalign*}

where $(d)$ and $(f)$ result from the relation that $\sigma'=\sigma + qLde_i$, $(e)$ and $(g)$ results from the definition of the oracle $\bO_{\epsilon}$. Comparing the RHS of $(g)$ and the RHS of $(c)$ gives us   
\begin{flalign} \label{eq:sigma-sigma'-compare-1}
    x_{t+h,i}(\sigma') - x_{t,i}(\sigma) \geq -\frac{hp}{q}|x_{t,i}(\sigma)-x_{t+h,i}(\sigma)|- \frac{ 3 \epsilon}{qLN}. 
\end{flalign}
Similarly, we have
\begin{flalign*}
     & F^t(x_t(\sigma'))-\langle \sigma',x_t(\sigma') \rangle + f^{t:t+h}(x_t(\sigma)) \\
    \stackrel{(h)}{\leq} & 
    F^t(x_{t+h}(\sigma))-\langle \sigma',x_{t+h}(\sigma) \rangle + f^{t:t+h}(x_t(\sigma)) + \epsilon \\
    \stackrel{(i)}{\leq} & 
    F^t(x_{t+h}(\sigma))-\langle \sigma',x_{t+h}(\sigma) \rangle + f^{t:t+h}(x_{t+h}(\sigma)) + Lh\|x_t(\sigma) - x_{t+h}(\sigma)\|+ \epsilon \\
    \stackrel{(j)}{\leq} & 
    F^t(x_{t+h}(\sigma))-\langle \sigma',x_{t+h}(\sigma) \rangle + f^{t:t+h}(x_{t+h}(\sigma)) + hpNL|x_{t,i}(\sigma)-x_{t+h,i}(\sigma)|+ \epsilon 
\end{flalign*}
here inequality $(h)$ result from the definition of the oracle $\bO_{\epsilon}$, $(i)$ results from the Lipschitz property of $f^\tau$ and $(j)$ results from (\ref{eq:norm_bd_assum}). On the other hand, we have 
\begin{flalign*}
     & F^t(x_t(\sigma'))-\langle \sigma',x_t(\sigma') \rangle + f^{t:t+h}(x_t(\sigma)) \\
    \stackrel{(k)}{=} &  F^t(x_t(\sigma'))-\langle \sigma,x_t(\sigma') \rangle + f^{t:t+h}(x^t(\sigma)) - qLN x_{t,i}(\sigma') \\
    \stackrel{(l)}{\geq} &  F^{t}(x_t(\sigma))-\langle \sigma,x_t(\sigma) \rangle + f^{t:t+h}(x^t(\sigma)) - qLN x_{t,i}(\sigma') - \epsilon \\
    \stackrel{(m)}{\geq} & F^{t+h}(x_{t+h}(\sigma))-\langle \sigma,x_{t+h}(\sigma) \rangle + qLN x_{t,i}(\sigma) - 2\epsilon  \\
    \stackrel{(n)}{=} & F^{t+h}(x_{t+h}(\sigma))-\langle \sigma',x_{t+h}(\sigma) \rangle + qLN \big(x_{t+h,i}(\sigma)-x_{t,i}(\sigma') \big) - 2\epsilon
\end{flalign*}
where $(k)$ and $(n)$ result from the relation that $\sigma'=\sigma + qLNe_i$, $(l)$ and $(m)$ results from the definition of the oracle $\bO_{\epsilon}$. Comparing the RHS of $(j)$ and the RHS of $(n)$ gives us   
\begin{flalign}
    x_{t,i}(\sigma') - x_{t+h,i}(\sigma) \geq -\frac{hp}{q}|x_{t,i}(\sigma)-x_{t+h,i}(\sigma)|- \frac{3 \epsilon}{qLN}. 
\end{flalign}
Finally, noting that $\epsilon>0$ and employing Lemma \ref{lem:monotonicity-homogenous} while setting $c=qLN$ we have
\begin{flalign}
    x_{t,i}(\sigma') - x_{t,i}(\sigma) &\geq -\frac{hp}{q}|x_{t,i}(\sigma)-x_{t+h,i}(\sigma)|- \frac{3 \epsilon}{qLN}, \label{eq:sigma-sigma'-compare-3} \\
    x_{t+h,i}(\sigma') - x_{t+h,i}(\sigma) &\geq -\frac{hp}{q}|x_{t,i}(\sigma)-x_{t+h,i}(\sigma)|- \frac{3 \epsilon}{qLN}. \label{eq:sigma-sigma'-compare-4}  
\end{flalign}
The result follows from combining (\ref{eq:sigma-sigma'-compare-1}-\ref{eq:sigma-sigma'-compare-4}). \hfill $\square$

We are now return to the proof of Lemma \ref{lem:generalized-stability-bd}. We being by noting that
\begin{flalign}
    \bE[\|x_{t+h}(\sigma)-x_{t}(\sigma)\|_1] = \sum_{i=1}^N \bE[|x_{t+h,i}(\sigma)-x_{t,i}(\sigma)|] = \sum_{i=1}^N \bE\Big[\bE\big[|x_{t+h,i}-x_{t,i}|\big|\sigma_{-i}\big]\Big] \label{eq:expand-norm}
\end{flalign}
where the last equality stems from employing the tower property of expectations. We will define the notation $\bE_{-i}[\cdot]$ to denote the conditional expectation $\bE[\cdot|\sigma_{-i}]$. Let us also define $\bar{x}_i(\sigma) = \max\{x_{t,i}(\sigma),x_{t+h,i}\}$ and $\underbar{x}_i(\sigma) = \min\{x_{t,i}(\sigma),x_{t+h,i}\}$. And finally, with the aim of employing Lemma \ref{lem:monotonicity-heterogenous}, let us define the event
\[\cE = \big\{\sigma \in \bR_{\geq0}^n \big| \|x_t(\sigma)-x_{t+h}(\sigma)\| \leq pN|x_{t,i}(\sigma)-x_{t+h,i}(\sigma)| \big\}\]

We are now interested in obtaining a $\sigma_{-i}$ independent bound on $\bE_{-i}[|x_{t+h,i}-x_{t,i}|]$ as then we will have the bound we are interested in. Note that $\bE_{-i}[|x_{t+h,i}-x_{t,i}|] = \bE_{-i}[\bar x_i(\sigma)] - \bE_{-i}[\underbar x_i(\sigma)]$ and motivated by this let us consider   
\begin{flalign}
    \bE_{-i}[\underbar x_i(\sigma)] & = \bP(\sigma_i< qLN) \bE_{-i}[\underbar x_i(\sigma)|\sigma_i< qLN] + \bP(\sigma_i\geq qLN)\bE_{-i}[\underbar x_i(\sigma)|\sigma_i\geq  qLN] \notag\\
    & \geq (1-\exp(-\nu qLN)) \big(\bE_{-i}[\bar x_i(\sigma)]-D\big) +\underbrace{\int_{qLN}^\infty \underbar x_i(\sigma)\nu \exp(-\nu \sigma_i) d\sigma_i}_{I_1} \label{eq:x_min_bd_intermidiate}
\end{flalign}
where the inequality above comes from noting that 
$\bE_{-i}[\bar x(\sigma)]$ and $\bE_{-i}[\underbar x(\sigma)|\sigma_i< qLd]$ are both elements of the convex set $\cX$ and the difference between their coordinates is bounded by the $L_1$-diameter $D$. Performing a change of variable $\sigma' = \sigma - qLN e_i$ we can rewrite integral $I_1$ as
\begin{flalign*}
    & \int_{qLN}^\infty \underbar x_i(\sigma)\nu \exp(-\nu \sigma_i) d\sigma_i  \\
    = & \exp{(-\nu qLN)} \int_{0}^\infty \underbar x_i(\sigma'+qLNe_i)\nu \exp(-\nu \sigma_i')  d\sigma_i' \\
    = & \exp{(-\nu qLN)}  \bE_{-i}\big[\underbar x_i(\sigma+qLNe_i)\big] \\
    = &\exp{(-\nu qLN)}  \bP(\cE|\sigma_{-i}) \bE_{-i}\big[\underbar x_i(\sigma+qLNe_i)|\cE\big] \\
    & + \exp{(-\nu qLN)}  \bP(\cE^c|\sigma_{-i}) \bE_{-i}\big[\underbar x_i(\sigma+qLNe_i)|\cE^c\big] \\
    \stackrel{(a)}{\geq} & \exp{(-\nu qLN)}  \bP(\cE|\sigma_{-i}) \bE_{-i}\bigg[\bar x_{i}(\sigma)-\frac{ph}{q}|x_{t,i}(\sigma)-x_{t+h,i}(\sigma)| - \frac{3\epsilon}{q LN}\bigg|\cE\bigg] \\
    & + \exp{(-\nu qLN)}  \bP(\cE^c|\sigma_{-i}) \bE_{-i}\bigg[\underbar x_i(\sigma) - \frac{2\epsilon}{qLN}\bigg|\cE^c\bigg] 
\end{flalign*}
where inequality $(a)$ is obtained by noting the definition of event $\cE$ and applying Lemma \ref{lem:monotonicity-heterogenous} and Lemma \ref{lem:monotonicity-homogenous}. Now note that when $\sigma \in \cE^c$ we have \[\|x_t(\sigma)-x_{t+h}(\sigma)\| \geq pN|x_{t,i}(\sigma)-x_{t+h,i}(\sigma)| = pN\big(\bar x_i(\sigma) - \underbar x_i(\sigma)\big).\]
This allows us to further lower bound the term on the RHS of $(a)$ as 
\begin{flalign*}
    & \int_{qLd}^\infty \underbar x_i(\sigma)\nu \exp(-\nu \sigma_i) d\sigma_i  \\
    \geq & \exp{(-\nu qLd)}  \bP(\cE|\sigma_{-i}) \bE_{-i}\bigg[\bar x_{i}(\sigma)-\frac{ph}{q}|x_{t,i}(\sigma)-x_{t+h,i}(\sigma)| - \frac{3\epsilon}{q Ld} \bigg|\cE\bigg] \\
    & + \exp{(-\nu qLd)}  \bP(\cE^c|\sigma_{-i}) \bE_{-i}\bigg[\bar x_i(\sigma) - \frac{1}{pd} \|x_t(\sigma)-x_{t+h}(\sigma)\| - \frac{3\epsilon}{qLd}\bigg|\cE^c\bigg] \\
     \geq & \exp(-\nu qLd) \bigg( \bE_{-i}[\bar x_i(\sigma)] -  \frac{3\epsilon}{qLd} - \frac{ph}{q} \bE_{-i}\big[|x_{t,i}(\sigma)-x_{t+h,i}(\sigma)|\big] \\ 
     & - \frac{1}{pd} \bE_{-i}\big[\|x_t(\sigma)-x_{t+h}(\sigma)\| \big]\bigg)
\end{flalign*}
where the final inequality stems simply from the fact that $\bP(\cE),\bP(\cE^c) \leq 1$. Substituting this lower bound for $I_1$ back in (\ref{eq:x_min_bd_intermidiate}) we have
\begin{flalign*}
    \bE_{-i}[\underbar x_i(\sigma)]  \geq &  \bE_{-i}[\bar x_i(\sigma)] - (1-\exp(-\nu qLN))D 
 - \exp(-\nu qLN) \bigg(  \frac{3\epsilon}{qLN} \\ 
     & + \frac{ph}{q} \bE_{-i}\big[|x_{t,i}(\sigma)-x_{t+h,i}(\sigma)|\big] 
     + \frac{1}{pN} \bE_{-i}\big[\|x_t(\sigma)-x_{t+h}(\sigma)\| \big]\bigg) \\
     \geq &  \bE_{-i}[\bar x_i(\sigma)] - \nu q L ND 
 - \bigg( \frac{3\epsilon}{qLN} + \frac{ph}{q} \bE_{-i}\big[|x_{t,i}(\sigma)-x_{t+h,i}(\sigma)|\big] \\
 & + \frac{1}{pN} \bE_{-i}\big[\|x_t(\sigma)-x_{t+h}(\sigma)\| \big]\bigg) 
\end{flalign*}
where the final equality results from $1-e^{-x} \leq x$ and from $e^x \leq 1$ for $x\leq 0$. Rearranging terms above and recalling $\bar x_i - \underbar x_i = |x_{t,i} - x_{t+1,i}|$ we get 
\begin{flalign*}
    \bigg(1-\frac{ph}{q}\bigg) \bE_{-i}\big[|x_{t,i}(\sigma)-x_{t+h,i}(\sigma)|\big] \leq \nu qLND +  \frac{3\epsilon}{qLN} + \frac{1}{pN} \bE_{-i}\big[\|x_t(\sigma)-x_{t+h}(\sigma)\| \big] 
\end{flalign*}
Recall that $p$ and $q$ are free variables in Lemmas \ref{lem:monotonicity-homogenous} and \ref{lem:monotonicity-heterogenous} and let us now assume $q > ph$.
By monotonicity of expectation and using (\ref{eq:expand-norm}) we get 
\begin{flalign*}
    & \bE[\|x_{t+h}(\sigma)-x_{t}(\sigma)\|] \leq \frac{q}{q-ph} \sum_{i=1}^d \bigg(\nu qLND + \frac{3\epsilon}{qLN} + \frac{1}{pN} \bE\big[\|x_t(\sigma)-x_{t+h}(\sigma)\| \big] \bigg) \\
\implies  &  \bigg(1-\frac{q}{p(q-ph)} \bigg)\bE[\|x_{t+h}(\sigma)-x_{t}(\sigma)\|]  \leq  \frac{q}{q-ph} \bigg(\nu qLN^2D + \frac{3\epsilon}{qL}\bigg)
\end{flalign*}
Finally, if we further assume $p>1$ and $q>\frac{p^2h}{p-1}$ we get 
\begin{flalign*}
    \bE[\|x_{t+h}(\sigma)-x_{t}(\sigma)\|]  \leq  \frac{qp}{(p-1)q-p^2h} \bigg(\nu qLN^2D + \frac{3\epsilon}{qL}\bigg) 
\end{flalign*}

Note that we need $p>1$ and $q >\max\{ph,\frac{p^2 h}{p-1}\}$. Setting $p=2$ and $q=5h$ gives us our result. \hfill $\square$

\subsection{Theorem \ref{thm:main-regret-bd-w-mem} statement and proof}
We will now restate the theorem with some additional details added in and then provide the complete proof for it. 
\begin{thm*} 
    Let $\sigma = [\sigma_1,\dots,\sigma_N]$ denote an exponential random vector such that $\sigma_n \stackrel{i.i.d}{\sim} \exp{(\nu)}$ for all $n\in [N]$. Let $\rho(\sigma,x) \triangleq - \langle x, \sigma \rangle$.
    Then, setting $\nu = \nicefrac{1}{\|U\|_1 (1+L) \sqrt{50 N (\Theta_H+H)}}$ and picking decisions according to 
    \begin{equation}
        \forall t>1, x^{t} = \bO_{\epsilon}\Big(-\langle \cY_{QR}(\cdot)^\intercal U^\intercal \cdot, G^{t-1} \rangle + \rho(\sigma,\cdot),\bbDelta_N\Big) \text{ and } x^1 = \bO_{\epsilon}\big( \rho(\sigma,\cdot),\bbDelta_N\big) \label{eq:FTPL-memory}
    \end{equation}
    allows us to bound the expected value of regret in (\ref{eq:regret-memory}) as 
    \begin{flalign*}
        \bE[Regret_M(H)] \leq 10 N \|U\|_1 (1+L) \sqrt{2 N \big( H + \Theta_H \big)} + \epsilon \big( 13 H + 1\big)
    \end{flalign*}
    where the expectation is taken over the distribution of $\sigma$. 
\end{thm*}
\textbf{Proof:} To avoid cluttered notation, throughout this proof we will drop the $QR$ subscript from the matrix response function and assume that $\cY$ captures the quantal response of followers. We begin by adding and subtracting terms in $Regret_M(H)$ as 
\begin{flalign*}
    \bE[Regret_M(H)]  = & \underbrace{max_{x\in \bbDelta_N} \langle \cY^\intercal(x) U^\intercal x, G^H\rangle - \sum_{t=1}^H \bE\big[\langle \cY(x^t)^\intercal U^\intercal x^t, g^t \rangle \big]}_{(A)} \\
     &+ \sum_{t=1}^H \underbrace{\bE\bigg[\Big\langle \big(\cY(x^t)-\cY(z^t)\big)^\intercal U^\intercal x^t, g^t \Big\rangle\bigg]}_{(B)}
\end{flalign*}


Let us first focus on terms of the form in $(B)$. Employing Corollary \ref{cor:Lip_leader_util}, noting that $\|g^t\|_1=1$ and recalling that $\|\cdot\|_\infty \leq \|\cdot\|_1$ for finite dimensional vectors we have 
\begin{flalign}
    \bE\bigg[\Big\langle \big(\cY(x^t)-\cY(z^t)\big)^\intercal U^\intercal x^t, g^t \Big\rangle\bigg] \leq \bE\big[L_u'\|x^t-z^t\|_1\big] \leq \frac{1}{b_t} \sum_{\tau=1}^{t-1} a_{t-\tau} \bE\big[L_u'\|x^t-x^\tau\|_1\big]. \label{eq:term_B_bd} 
\end{flalign}
where the last inequality results from (\ref{eq:avg_leader-strat}) and by applying triangle inequality. Now shifting focus to term (A), we observe that it is identical to the regret (\ref{eq:regret-no-memory}) of the learner when facing memory-less followers. Once again, we note that (\ref{eq:FTPL-memory}) is equivalent to (\ref{eq:approx-FTPL}) when $f^t = -\langle \cY(x)^\intercal U^\intercal x, g^t \rangle$ and $\cX = \bbDelta^N$. Then, invoking Theorem \ref{thm:FTPL-general-bound} gives us 
\begin{flalign*}
         (A) \leq & \sum_{t=1}^H \bE\big[\langle \cY^\intercal(x^{t+1}) U^\intercal x^{t+1}, g^t \rangle\big] - \bE\big[\langle \cY^\intercal(x^t) U^\intercal x^t, g^t \rangle\big] + \bE\big[\langle x^1 - x' ,\sigma \rangle\big] + \epsilon (H+1) \\
         \stackrel{(a)}{\leq} & \sum_{t=1}^H L_u\underbrace{\bE\big[\|x^t-x^{t+1}\|_1\big]}_{(C)} + \underbrace{\bE\big[\|x^1 - x'\|_\infty \|\sigma\|_1 \big]}_{(D)} + \epsilon (H+1) 
    \end{flalign*}
where the inequality in $(a)$ follows from Corollary \ref{cor:Lip_leader_util} and from Holder's inequality.   
Noting that $x^1$ and $x'$ are both elements of $\bbDelta^N$ and by properties of exponential random variables we can bound term $D$ as
\[(D) \leq \sum_{n\in [N]} \bE\big[ \sigma_n \big] = \frac{N}{\nu}.\] 
We will now invoke Lemma \ref{lem:generalized-stability-bd} to bound the terms in the RHS of (\ref{eq:term_B_bd}) and terms of the form in (C). To do so, we note that (\ref{eq:FTPL-memory}) is equivalent to (\ref{eq:non-cvx-learn-opt-oracle}) when $f^t = -\langle \cY(x)^\intercal U^\intercal x, g^t \rangle$ and $\cX = \bbDelta^N$. Note, also that the values of $D$ and $L$ in Lemma \ref{lem:generalized-stability-bd} are $1$ and $L_u$ respectively. So we have
\begin{flalign*}
    \bE[Regret_M(H)]  \leq & \sum_{t=1}^H  \bigg(L_u\bE\big[\|x^t-x^{t+1}\|\big] +\frac{1}{b_t} \sum_{\tau=1}^{t-1} a_{t-\tau} \bE\big[L_u'\|x^t-x^\tau\|\big]\bigg) + \frac{N}{\nu} + \epsilon (H+1)\\
    \leq & \sum_{t=1}^H  \Bigg(50\nu L_u^2 N^2 + 6\epsilon +\frac{L_u'}{b_t} \sum_{\tau=1}^{t-1} a_{t-\tau} \bigg(50\nu (t-\tau) L_uN^2 + \frac{6\epsilon}{(t-\tau) L_u}\bigg)\Bigg) \\
    & + \frac{N}{\nu} + \epsilon (H+1) \\ 
    \leq & \sum_{t=1}^H  \Bigg(50\nu L_u^2 N^2 + 6\epsilon +L_u'  \bigg(50\nu \theta_t L_uN^2 + \frac{6\epsilon}{L_u}\bigg)\Bigg) + \frac{N}{\nu} + \epsilon (H+1) \\ 
\end{flalign*}

where the final inequality results from noting that $\frac{1}{b_t} \sum_{\tau=1}^{t-1} \frac{a^{t-\tau}}{(t-\tau)} \leq \frac{1}{b_t} \sum_{\tau=1}^{t-1} a^{t-\tau} \leq 1$ and by defining $\theta_t \triangleq \frac{1}{b_t} \sum_{\tau=1}^{t} a^{t-\tau} (t-\tau)$. 
Noting that $\Theta_H = \sum_1^H \theta_t$ we get our bound on the expected regret as
\begin{flalign*}
    \bE[Regret_M(H)]  \leq &  50\nu N^2 \big(L_u^2 H + \Theta_H L_u L_u' \big)+ \epsilon  \bigg(7H+6\frac{L_u'}{L_u}H+1 \bigg) + \frac{N}{\nu} 
\end{flalign*}
Finally, substituting the values of $L_u$ and $L_u'$ gives us 
\begin{flalign*}
    \bE[Regret_M(H)]  \leq & 50\nu N^2 \|U\|^2_1 \big((1+L)^2 H + \Theta_H L(1+L) \big) + \frac{N}{\nu} + \epsilon \bigg( 7H+6\frac{L_u'}{L_u} H + 1\bigg) \\
    \leq & 50\nu N^2 \|U\|^2_1 (1+L)^2 \big( H + \Theta_H \big) + \frac{N}{\nu} + \epsilon \big( 13 H + 1\big)
\end{flalign*}
Substituting the value for $\nu$ gives us the required bound.  
\hfill $\square$

\textbf{Proof of Corollary \ref{robustness}}:
It follows directly from futher upper-bounding the last term in the the proof above by $50\nu N^2 \|U\|^2_1 (1+L)^2 \big( H + \overline\Theta \big) + \frac{N}{\nu} + \epsilon \big( 13 H + 1\big)$ and substituting the appropriate $\nu$.

\section{Discussion on Claim \ref{cla:MK_LPs}} \label{app:discuss-MK-LP}

Let us first provide a justification for this claim. We are interested in an oracle that finds $\epsilon$-optimal solutions to problems of the form 
\begin{flalign}
    \max_{x\in \bbDelta_N}  \sum_{k\in [K]} (\sigma_k + G_k) x^\intercal U y^k_{BR}(x) = \max_{x\in \bbDelta_N}  \sum_{k\in [K]} G_k' x^\intercal U y^k_{BR}(x)\label{eq:LP-prob-state}
\end{flalign}
where $G_k'=G_k+\sigma_k$. 
Now, we note that the basis vector $e_i\in \bR^M$ is \emph{a} best response for a follower with type $k$ if and only if 
\[ x\in P_i^k \triangleq \big\{ z\in \bbDelta_M | \langle V_i^k,z\rangle \geq \langle V_j^k,z\rangle \quad \forall j \in [M]\setminus\{i\} \big\}.\]
Note here that $P_i^k$ is a convex polytope as it is generated by the intersection of finitely many halfspaces. Since $\{P_i^k\}_{i}$ covers $\bbDelta_N$, when evaluating the leader's utility for any strategy $x\in \bbDelta_N$, it suffices to consider that each follower picks responses solely from the set of basis vectors $\cE = \{e_1,\dots,e_M\}$ . When faced with only a single follower of type $k$, this observation allows the leader to efficiently compute her optimal strategy by solving $M$ linear programs (LPs) \cite{conitzer2006computing} of the form
\[\max_{x\in \bbDelta_N \cap P_i^k} x^\intercal U e_i \quad \forall i\in [M].\]
The leader then can pick her optimal strategy as the optimum corresponding to the LP that achieves the highest value. We now take a similar approach in computing the solution to (\ref{eq:LP-prob-state}).

Let $i_k$ denote the action picked by the $k^{th}$ follower. Then, every strategy profile of the followers can be denoted as $\{e_{i_k}\}_{k\in [K]}\in \cE^K$. By restricting ourselves to one such strategy profile we can frame the following LP:
\begin{subequations} \label{eq:LP_for_oracle}
\begin{flalign}
& \max_{x\in \bbDelta_N \cap \big(\bigcap_{k=1}^K P_{i_k}^k\big)}  \sum_{k\in [K]} G_k' x^\intercal U e_{i_k} \notag\\
= & \max_{x\in \bbDelta_N}   \bigg\langle x,  \sum_{k\in [K]} G_k' U_{i_k} \bigg\rangle \\
& S.T \quad  \langle V_{i_k}^k,z\rangle \geq \langle V_j^k,z\rangle \quad \forall j \in [M]\setminus\{i_k\}, \forall k\in[K] \label{eq:LP-constraints} 
\end{flalign}
\end{subequations}
The solution to (\ref{eq:LP-prob-state}) is then the optimum corresponding to the LP (of the form in (\ref{eq:LP_for_oracle})) that achieves the highest value across the strategy profiles. However, noting that the cardinality of $\cE^K$ is $M^K$, this requires us to solve an exponential number of LPs. Note that each LP of the form in (\ref{eq:LP_for_oracle}) can be solved is polynomial time as it only has polynomially many ($M-1\times K$) constraints.

Finally, we note that the solutions $x^*$ picked according to (\ref{eq:LP_for_oracle}) may satisfy equality in some of the constraints in (\ref{eq:LP-constraints}). This poses a challenge because in such cases, a follower might be indifferent between two or more options for their best response. Consequently, we cannot guarantee that a follower of type $k$ will select $e_{i_k}$ when the leader plays $x^*$. However, this issue can be easily resolved by selecting a point $x'$ within the interior of the set $\bigcap_{k=1}^K P_{i_k}^k$ that is "close" to $x^*$. Specifically, choosing a point such that $\|x'-x^*\|_1\leq \delta$, with $\delta = \epsilon\big(\sum_k |G_k'|\overline{U}\big)^{-1}$, is adequate to ensure $\epsilon$-optimality of $x'$ while also guaranteeing unique follower responses.


\paragraph{Connection to the approach by \citet{balcan2015commitment}}: 
Noting that followers play pure strategies when best responding, \citet{balcan2015commitment} consider a partitioning of the leader's strategy space according to regions where followers' strategy profile is constant. Indeed the elements of this partition are nothing but the convex polytopes $\big\{\bbDelta_N \cap \big(\bigcup_{k=1}^K P_{i_k}^k\big)\big|i_k\in[M]^K\big\}$ that appear as the feasible set in (\ref{eq:LP_for_oracle}). 

Then, noting that in each of these regions the leader has a linear utility function, they restrict the set of leader strategies that need to be considered to the vertices of these regions. Note that this observation holds irrespective of the values taken by $G_k'$ in (\ref{eq:LP_for_oracle}). Consequently, they are left with exponentially many leader strategies to consider in each region. They pre-compute these strategies for each of the $M^K$ regions and then employ a Prediction-from-Expert-Advice-style online learning algorithms to pick the best strategy from these. This involves computing the payoff of each of these exponential many strategies in every step of the learning horizon.

Rather than store every vertex of this region, with the knowledge of $G_k'$, our algorithm merely computes the vertex (strategy) where the leader enjoys the highest payoff through an LP. However, we are still left with solving $M^K$ such LPs to then find the best strategy across all regions. Thus, the two approaches are intimately connected and are both computational expensive.

\section{Simulation results} \label{app:sims}

\begin{figure}
    \centering
    \begin{subfigure}[b]{0.49\textwidth}
        \centering
        \includegraphics[width=\textwidth]{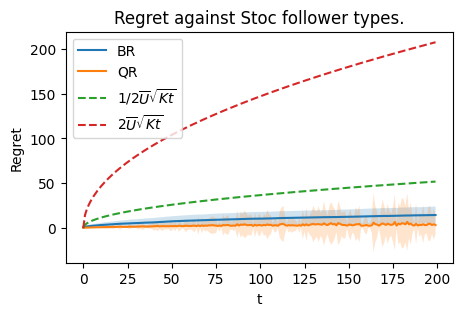}  
        \caption{Incurred regret compared to theoretical bounds.}
        \label{fig:no-mem-stoc}
    \end{subfigure}
    \hfill
    \begin{subfigure}[b]{0.49\textwidth}
        \centering
        \includegraphics[width=\textwidth]{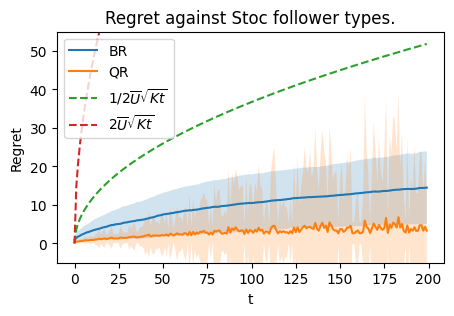}
        \caption{Zoomed in view of Figure \ref{fig:no-mem-stoc}.}
        \label{fig:no-mem-stoc-zoomed}
    \end{subfigure}
    \hfill
    \begin{subfigure}[b]{0.49\textwidth}
        \centering
        \includegraphics[width=\textwidth]{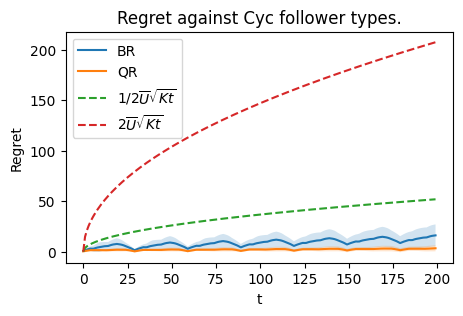} 
        \caption{Incurred regret compared to theoretical bounds.}
        \label{fig:no-mem-cyc}
    \end{subfigure}
    \hfill
    \begin{subfigure}[b]{0.49\textwidth}
        \centering
        \includegraphics[width=\textwidth]{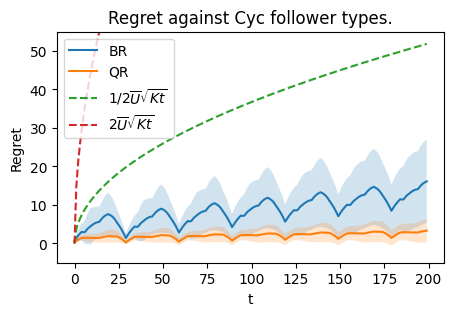}
        \caption{Zoomed in view of Figure \ref{fig:no-mem-cyc}.}
        \label{fig:no-mem-cyc-zoomed}
    \end{subfigure}
    \caption{\textbf{(Experimental results for followers without memory.)} 
     Figure \ref{fig:no-mem-stoc}  (zoomed in \ref{fig:no-mem-stoc-zoomed}) showcases the performance of our Algorithm \ref{alg:FTPL_no_mem} when facing a sequence of followers that are picked uniformly at random \textbf{(Stoc)}.  Figure \ref{fig:no-mem-cyc} (zoomed in \ref{fig:no-mem-cyc-zoomed}) showcases its performance when facing a sequence of followers that are picked in a round robin fashion \textbf{(Cyc)}.  In both situations, it took $\sim 3500$ secs to run this simulation for $S=400$ iterations.}
    \label{fig:regret_no-mem}
\end{figure}

We now present some simulation results where we apply the aapproaches presented in this paper to learn from synthetically generated data. Let us first present the two algorithms described in Theorem \ref{thm:main-regret-bd-no-mem} (Algorithm \ref{alg:FTPL_no_mem}) and Theorem \ref{thm:main-regret-bd-w-mem} (Algorithm \ref{alg:FTPL_mem}) succintly. 
\begin{algorithm}
\caption{FTPL for Repeated SGs (Followers without Memory)}
\begin{algorithmic}[1]
\REQUIRE $H$, $\nu > 0$, $\cY$, $U$
\STATE Draw perturbation $\sigma_i \stackrel{i.i.d}{\sim} U[0,2\nu^{-1}],\; i\in [K]$
\STATE Set $G^0 = [0,\dots,0] \in \bR^K$
\FOR{each round $t=1,2,\ldots,H$}
    \STATE Pick leader $x^t = \bO_\epsilon\big(-\langle\cY(\cdot)^\intercal U^\intercal \cdot,G^{t-1}+\sigma\rangle\big)$
    \STATE Observe follower type $g^t$; Receive Payoff $\langle\cY(x^t)^\intercal U^\intercal x^t ,g^{t}\rangle$ 
    \STATE Update $G^{t}=G^{t-1}+g^t$
\ENDFOR
\end{algorithmic}
\label{alg:FTPL_no_mem}
\end{algorithm}

\begin{algorithm}
\caption{FTPL for Repeated SGs (Followers with Memory)}
\begin{algorithmic}[1]
\REQUIRE $H$,$\nu > 0$, $\cY_{QR}$,$U$,$\{a_t\}_{1}^H$
\STATE Draw perturbation $\sigma_i \stackrel{i.i.d}{\sim} \exp{(\nu)},\; i\in [N]$
\STATE Set $G^0 = [0,\dots,0] \in \bR^K$
\FOR{each round $t=1,2,\ldots,H$}
    \STATE Pick leader $x^t = \bO_\epsilon\big(-\langle\cY_{QR}(\cdot)^\intercal U^\intercal \cdot,G^{t-1}\rangle +\langle \sigma,\cdot \rangle\big)$
    \STATE Update average $z^t$ according to (\ref{eq:avg_leader-strat}) 
    \STATE Observe follower type $g^t$; Receive Payoff $\langle\cY_{QR}(z^t)^\intercal U^\intercal x^t ,g^{t}\rangle$ 
    \STATE Update $G^{t}=G^{t-1}+g^t$
\ENDFOR
\end{algorithmic}
\label{alg:FTPL_mem}
\end{algorithm}

\subsection{Methods}
For the simulations, we synthetically generate two types of sequences of followers $\{g^t\}_{t=1}^T$. 
\begin{enumerate}
    \item \textit{Stochastic Sequence} \textbf{(Stoc)} - In every round, the type of the follower is picked independently and uniformly from $[K]$ i.e. $g^t \stackrel{i.i.d}{\sim} Uniform(\cE)$ for all $t$, where $\cE$ is the set of basis vectors in $\bR^K$. 
    \item \textit{Round Robin Sequence} \textbf{(Cyc)} - We pick every type from the set $[K]$ in a cyclic fashion. After picking a type, it is held constant for a fixed period $L$ before moving to the next type. In other words, $g^t = e_i$ if $(\lfloor \nicefrac{t}{L}\rfloor\bmod K) + 1= i$.     
\end{enumerate}
When considering followers with memory we will consider the following two models: 
\begin{enumerate}
    \item \textit{Finite Memory} \textbf{(FM)} - Followers have a bounded memory of length $B$ and they weigh all past decisions equally ($a_s=1$ if $s<B$ and $a_s=0$ otherwise).  
    \item \textit{Discount Weighted Memory} \textbf{(DM)} - Followers weigh past actions by a discount factor $0<\gamma<1$ (i.e. $a_s = \gamma^s$).
\end{enumerate}

\paragraph{Oracles} As highlighted in Section \ref{sec:comp_aspects}, the design of the oracle we use depends on the nature of the followers response. For best-responding followers, the design of our oracle is discussed in Appendix \ref{app:discuss-MK-LP}. In addition, we use one additional heuristic to speed up computation. Note that of the $M^K$ linear programs (Claim \ref{cla:MK_LPs}) we need to solve, it is possible that some of them are infeasible. Moreover, the feasibility depends solely on the constraints in (\ref{eq:LP-constraints}) and not on either the perturbation $\sigma$ we employ or the sequence of follower types $\{g^t\}_{t}$. That allows us to pre-compute the feasibility of these LP and eliminate the ones that are infeasible. We observe that for the simulations we run, this speeds up our computation by a factor of $\sim 20$. For quantal responding followers, we employ the Simplicial Homology Global Optimization (SHGO) non-linear solver that is implemented in the \href{https://docs.scipy.org/doc/scipy/reference/generated/scipy.optimize.shgo.html}{SciPy} package as \lstinline{scipy.optimize.shog}. All simulation runs employ default values for the optional parameters in \lstinline{scipy.optimize.shog}. 

\textbf{Simulation Parameters:} We run all our simulations over a learning horizon of length $H=200$. We pick $N=M=3$ and $K=6$. When considering quantal responding followers we consider $\eta=2$ based on observed experimental values \cite{mckelvey1995quantal}. When considering (\textbf{Cyc}) cost functions we take $L=5$. When considering followers with memory, we take $B=10$ and $\gamma = 0.9$. 
All simulations were performed on a personal computer with CPUs (Intel Core i7-8550U CPU  $\times$ 8) and 16 GB RAM.  

When using randomly generated payoff matrices $U$ and $\{V^k\}_k$ for the players, we observed that the regret incurred by our algorithm remained very close to zero. In an attempt to capture instances closer to worst case scenario we picked
\[U = \begin{bmatrix}
    3 & 2 & 1 \\
    2 & 3 & 1 \\
    1 & 2 &3
\end{bmatrix}\]
and values for $V^k$ were obtained by permuting the columns of  $-\bI_{3\times 3}$ (giving us $K=6$ combinations).

\begin{figure}
    \centering
    \begin{subfigure}[b]{0.49\textwidth}
        \centering
        \includegraphics[width=\textwidth]{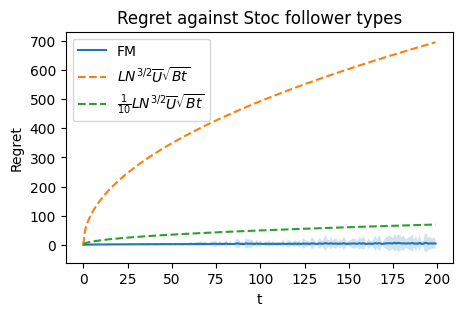}  
        \caption{Incurred regret compared to theoretical bounds.}
        \label{fig:bd-mem-stoc}
    \end{subfigure}
    \hfill
    \begin{subfigure}[b]{0.49\textwidth}
        \centering
        \includegraphics[width=\textwidth]{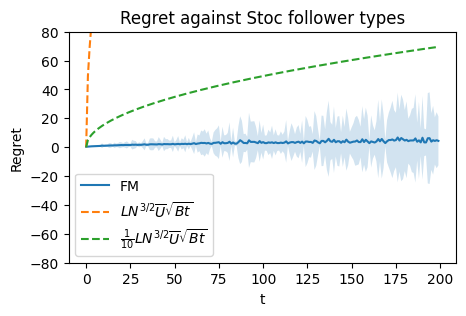}
        \caption{Zoomed in view of Figure \ref{fig:bd-mem-stoc}.}
        \label{fig:bd-mem-stoc-zoomed}
    \end{subfigure}
    \hfill
    \begin{subfigure}[b]{0.49\textwidth}
        \centering
        \includegraphics[width=\textwidth]{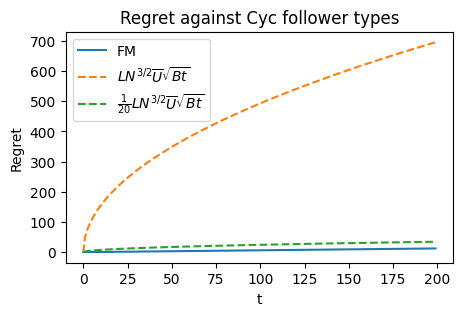} 
        \caption{Incurred regret compared to theoretical bounds.}
        \label{fig:bd-mem-cyc}
    \end{subfigure}
    \hfill
    \begin{subfigure}[b]{0.49\textwidth}
        \centering
        \includegraphics[width=\textwidth]{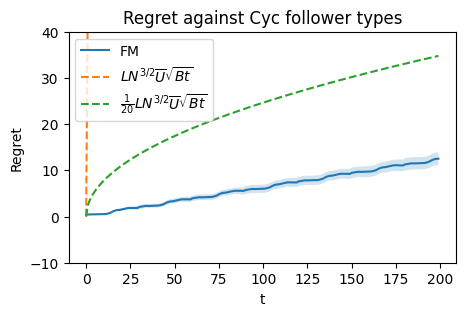}
        \caption{Zoomed in view of Figure \ref{fig:bd-mem-cyc}.}
        \label{fig:bd-mem-cyc-zoomed}
    \end{subfigure}
    \caption{\textbf{(Experimental results for followers with finite memory \textbf{(FM)}.)} 
     Figure \ref{fig:bd-mem-stoc}  (zoomed in \ref{fig:bd-mem-stoc-zoomed}) showcases the performance of our Algorithm \ref{alg:FTPL_mem} when facing a sequence of followers that are picked uniformly at random \textbf{(Stoc)}.  Figure \ref{fig:bd-mem-cyc} (zoomed in \ref{fig:bd-mem-cyc-zoomed}) showcases its performance when facing a sequence of followers that are picked in a round robin fashion \textbf{(Cyc)}. In both situations, it took $\sim 2000$ secs to run this simulation for $S=400$ iterations.}
    \label{fig:regret_bd-mem}
\end{figure}

\subsection{Results}
\begin{figure}
    \centering
    \begin{subfigure}[b]{0.49\textwidth}
        \centering
        \includegraphics[width=\textwidth]{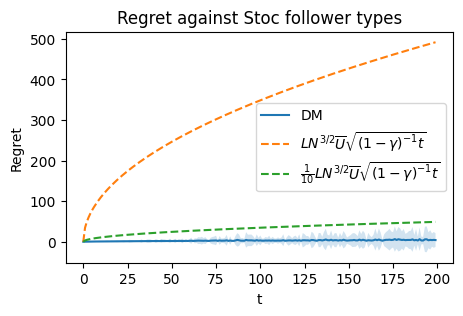}  
        \caption{Incurred regret compared to theoretical bounds.}
        \label{fig:disc-mem-stoc}
    \end{subfigure}
    \hfill
    \begin{subfigure}[b]{0.49\textwidth}
        \centering
        \includegraphics[width=\textwidth]{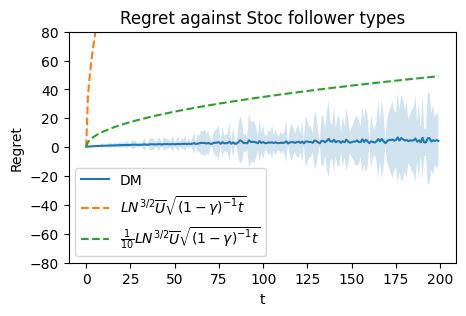}
        \caption{Zoomed in view of Figure \ref{fig:disc-mem-stoc}.}
        \label{fig:disc-mem-stoc-zoomed}
    \end{subfigure}
    \hfill
    \begin{subfigure}[b]{0.49\textwidth}
        \centering
        \includegraphics[width=\textwidth]{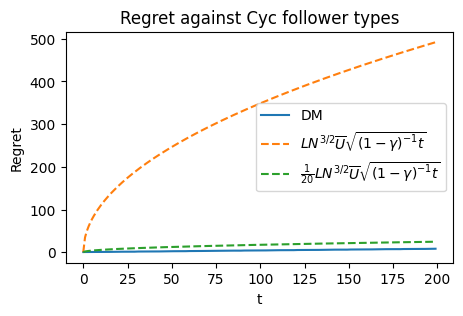} 
        \caption{Incurred regret compared to theoretical bounds.}
        \label{fig:disc-mem-cyc}
    \end{subfigure}
    \hfill
    \begin{subfigure}[b]{0.49\textwidth}
        \centering
        \includegraphics[width=\textwidth]{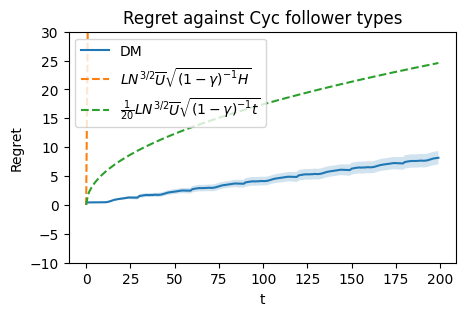}
        \caption{Zoomed in view of Figure \ref{fig:disc-mem-cyc}.}
        \label{fig:disc-mem-cyc-zoomed}
    \end{subfigure}
    \caption{\textbf{(Experimental results for followers with discount weighted memory \textbf{(DM)}.)} 
     Figure \ref{fig:bd-mem-stoc}  (zoomed in \ref{fig:bd-mem-stoc-zoomed}) showcases the performance of our Algorithm \ref{alg:FTPL_mem} when facing a sequence of followers that are picked uniformly at random \textbf{(Stoc)}.  Figure \ref{fig:bd-mem-cyc} (zoomed in \ref{fig:bd-mem-cyc-zoomed}) showcases its performance when facing a sequence of followers that are picked in a round robin fashion \textbf{(Cyc)}. In both situations, it took $\sim 2000$ secs to run this simulation for $S=400$ iterations.}
    \label{fig:regret_disc-mem}
\end{figure}

All results presented in Figures \ref{fig:regret_no-mem}, \ref{fig:regret_bd-mem} and \ref{fig:regret_disc-mem} have been generated by running $S=400$ iterations with a new sequence of followers $\{g^t\}_t$ generated every iteration. In every case, we indicate the average regret incurred by our algorithms across these $S$ iterations as well as the $1$-sigma bounds (shaded region) around this average. Additionally, in all the figures we plot the theoretical bounds presented in Theorems \ref{thm:main-regret-bd-no-mem} and \ref{thm:main-regret-bd-w-mem} as well as scaled down versions of the bound. 

The scaled version qualitatively indicates the gap between our theoretical bound and the average regret incurred in the instances we generate. Note that this gap can originate due to two reasons: (i) our algorithm runs particularly well on the instances we are considering, or (ii) the bounds we have obtained are loose.\footnote{Note that our bounds are tight in the horizon length $H$ as it is a well known that regret in online optimization is $\Omega{(H)}$ \cite{hazan2016introduction,cesa2006prediction}.} While it is difficult to generate instances that lead to the worst-case regret, we do observe that this gap is smaller in Figure \ref{fig:regret_no-mem} than in Figures \ref{fig:regret_bd-mem} and Figure \ref{fig:regret_disc-mem}. Noting that the algorithms we implement in the two situations are different, we hypothesize that this gap is the result of a weak bound on the Lipschitz constant in Claim \ref{claim:diff_y_bd}. Nevertheless, our simulation results validate the theoretical guarantees we propose (and prove) in Section \ref{sec:results}.



\end{document}